\documentclass{article}

\usepackage{arxiv}

\usepackage[utf8]{inputenc} 
\usepackage[T1]{fontenc}    
\usepackage{hyperref}       
\usepackage{url}            
\usepackage{booktabs}       
\usepackage{amsfonts}       
\usepackage{nicefrac}       
\usepackage{microtype}      
\usepackage{lipsum}

\usepackage{lineno,hyperref}
\usepackage{amsfonts}
\usepackage{amsmath}
\usepackage{amsthm}
\usepackage{cases}
\usepackage[shortlabels]{enumitem}

\usepackage{bbm}

\usepackage{tikz,pgf,pgfplots}

\usetikzlibrary{arrows,shapes,snakes,automata,backgrounds,petri,fit,intersections,quotes,positioning}

\usepackage{tikz-3dplot}

\pgfplotsset{compat=1.14}

\theoremstyle{plain}
\newtheorem{theorem}{Theorem}[section]

\newtheorem{proposition}[theorem]{Proposition}

\theoremstyle{definition}

\theoremstyle{remark}
\newtheorem{remark}[theorem]{Remark}

\definecolor{cof}{RGB}{219,144,71}
\definecolor{pur}{RGB}{186,146,162}
\definecolor{greeo}{RGB}{91,173,69}
\definecolor{greet}{RGB}{52,111,72}

\title{Personal Finance Decisions with Untruthful Advisors: an Agent-Based Model}

\author{
  Loretta Mastroeni
  Dept. of Economics\\
  Roma Tre University\\
  Via Silvio D'Amico 77, 00145 Rome, Italy \\
  \texttt{loretta.mastroeni@uniroma3.it} \\
   \And
  Maurizio Naldi \\
  Dept. of  Law, Economics, Politics and Modern languages\\
  LUMSA University\\
  Via Marcantonio Colonna 19, 00192 Rome, Italy \\
  \texttt{m.naldi@lumsa.it} \\
  \And
  Pierluigi Vellucci \\
  Dept. of Economics\\
  Roma Tre University\\
  Via Silvio D'Amico 77, 00145 Rome, Italy \\
  \texttt{pierluigi.vellucci@uniroma3.it} \\
}

\begin{document}
\maketitle

\begin{abstract}
Investors usually resort to financial advisors to improve their investment process until the point of complete delegation on investment decisions. Surely, financial advice is potentially a correcting factor in investment decisions but, in the past, the media and regulators blamed biased advisors for manipulating the expectations of naive investors. In order to give an analytic formulation of the problem, we present an Agent-Based Model formed by individual investors and a financial advisor. We parametrize the games by considering a compromise for the financial advisor (between a sufficient reward by bank and to keep his/her reputation), and a compromise for the customers (between the desired return and the proposed return by advisor). Then we obtain the Nash equilibria and the best response functions of the resulting game. We also describe the parameter regions in which these points result acceptable equilibria and the greediness/naivety of the customers emerge naturally from the model. Finally, we focus on the efficiency of the best Nash equilibrium.
\end{abstract}

\keywords{Opinion dynamics \and Agents-based model \and Personal finance \and Price of Stability}

\section{Introduction}
\label{intro}


Decisions concerning personal finance are taken by individuals on the basis of a variety of factors. For example, the investment decision process appears to incorporate a broad range of variables that may influence the individual investor's behaviour, such as the perceived ethics of a firm, and recommendations from individual stock brokers or friends/coworkers \cite{Nagy94}. Actually, investors usually resort to financial advisors to improve their investment process until the point of complete delegation on investment decisions. Surely, financial advice is potentially a correcting factor in investment decisions \cite{Fischer}. 


Nevertheless, in the aftermath of the recent financial bubbles, the media and the regulators usually placed much of the blame on biased advisors for manipulating the expectations of naive investors \cite{HONG2008268}. According to this view, an analyst may receive incentives to generate biased, optimistic forecasts while naive individual investors are unable to recognize that these biased recommendations are motivated by incentives to sell financial products. This means that, when asked for a professional advice (i.e. an opinion), advisors may not straightforwardly state what they truly think, but rather be tempted to misrepresent their opinion to conform to the bank they are paid by. In any case, the role of financial advisors, as well as that of other influencers, is to be properly accounted for in an analysis of personal finance decisions.

However, though the decisions taken in personal finance have been a subject of interest in a number of papers (see, e.g., \cite{BRUHN2011315,CHIANGLIN2008373,JENSEN201528,Konicz2015,kraft_steffensen_2008}), so far the adopted framework has considered an individual acting without interaction with influencers of any kind. 

In this paper, we wish instead to consider such interactions, with the aim of understanding how the opinions of an individual investor may change under the influence of his/her advisors, considering the aims of the stakeholders involved. For this purpose we formulate an Agent-Based Model (ABM) that includes three classes of agents: a bank, a financial advisor, a set of investors or customers. This model mimics the environment an individual investor finds when it manageshis/her investment through the local branch of a bank, where a financial advisor oversees a group of the bank's customers. The different aims of the stakeholders are recognized by resorting to a game-theoretic model, where  each personal investment/advise corresponds to selecting a strategy, and the agents’ payoff depends on the strategies chosen by him/herself and other players. We refer to this game as the \emph{personal finance game} and provide the following contributions:

\begin{itemize}
    \item We introduce an ABM to address the personal finance game (Section \ref{sec:model}). The major advantage that we expect from the adoption of an analytic framework like ABM is that analytic derivations of the properties of the model can be equally used as descriptive and as prescriptive tools, as widely noticed in the literature (e.g. \cite{MONICA20172272}, and also \cite{PARESCHI2017201,Vellucci2017,MasNaVel,MVNsurvey,MNVWOACEUR,MNVODS}).
    \item We obtain the best response functions and the Nash equilibria of the resulting game (Section \ref{sec:nash}). We also describe the parameter regions in which these points represent acceptable equilibria (Section \ref{sec:adm_nash}). Surprisingly,  without constraints on the returns (except for the fact that the proposed returns by advisor differ from the one desired by customers), the equilibrium is reached when the customer expects a return bigger than the one the advisor proposes instead.
    \item We provide a mathematical description of the boundary of the utility functions domain $\mathcal{D}$ (Section \ref{sec:boundary}).
    \item We introduce the social welfare in this context and obtain an analytic formulation of the Price of Stability (the generalization of the Price of Anarchy when more equilibria are present) for our ABM (Section \ref{sec:price}) thanks to the analytic formulation of $\mathcal{D}$ (given in the previous point).
\end{itemize}



\section{Related literature}

Our paper is related to the literature framework of \emph{opinion formation games} that relax the assumption of \emph{truthfulness} (a.k.a as honesty) in the process of opinion formation, allowing game players to express some opinions which need not coincide with their true opinions.

The players whose opinions we wish to model are represented as nodes of a social network (i.e., vertices on a graph), where the links between the nodes represent the direct influence between players in forming their opinions.

We therefore introduce a connected undirected graph $G =(\mathcal{V}, \mathcal{E})$ be  with $|\mathcal{V}| =n$ and for each edge $e =(i, j) \in \mathcal{E}$ let $w_{ij}\geq0$ be its weight. Let $W=[w_{ij}]_{ij}$ be the matrix of weights. Every vertex of the graph (i.e., each player or agent) is characterized by an internal opinion $s_i$ and a stated opinion $z_i$. The set of neighbors of agent $i$ in the social network represented by the graph $G$ is denoted by $N(i)$.

This game can be expressed as an instance $\left(G,W, \textbf{s}, \textbf{z} \right)$ that combines a weighted graph
$(G,W)$ and the vectors of opinions $\textbf{s}=(s_1,\dots,s_n)$ and $\textbf{z}=(z_1,\dots,z_n)$, which are attributes of the nodes. The internal opinion $s_i$ is unchanged and not affected by opinion updates, while each player's strategy is represented by his/her stated opinion $z_i$, which may be different from his/her $s_i$ and gets updated \cite{BINDEL2015248,Gionis,Bhawalkar:2013:COF:2488608.2488615,FERRAIOLI201696,CHIERICHETTI201811,auletta2016generalized,Auletta:2017:RDP:3091282.3091307,Bilo,CHEN2016808}. The paper by Buechel et al. \cite{BUECHEL2015240} differs from all these papers by mainly as it considers true and stated opinions evolving over time according to different laws. 

In \cite{BUECHEL2015240} the utility of agent $i$ depends on the distance of true opinion $s_i$ to stated opinion $z_i$ as well as on the distance of stated opinion $z_i$ to group opinion $q_i$. Bindel et al. \cite{BINDEL2015248} study the \emph{price of anarchy} --- the ratio between the cost of the Nash equilibrium and the cost of the optimal solution --- in a game of opinion formation. They assume that person $i$ has an internal opinion $s_i$, which remains unchanged from external influences, and a stated opinion $z_i$ which is updated as a weighted sum of his/her' neiughbours' stated opinions
\begin{equation}
\label{eq:zi_bindel}
    z_i=\frac{s_i+\sum_{j\in N(i)} w_{ij}z_j}{1+\sum_{j\in N(i)} w_{ij}}
\end{equation}
where $w_{i,j}\geq 0$. Both opinions are assumed to be real numbers. Updating $z_i$ as in (\ref{eq:zi_bindel}) allows to minimize the cost function
\begin{equation}
\label{eq:bindel_cost}
(z_{i}-s_i)^2 + \sum_{j\in N(i)} w_{ij} (z_{j} - z_{i})^2\, .
\end{equation}
Both papers, \cite{BINDEL2015248} and \cite{BUECHEL2015240}, are inspired by classical models due to DeGroot \cite{Degroot} and Friedkin-Johnsen \cite{Friedkin}. Also in \cite{BUECHEL2015240} both opinions $z_i$ and $s_i$ are assumed to be real numbers.

Gionis et al. \cite{Gionis} follow the framework of Bindel et al. \cite{BINDEL2015248}, by considering equations (\ref{eq:zi_bindel}) and (\ref{eq:bindel_cost}) as update rule and personal cost function. The internal and external opinions have been modeled as real values in the interval $[0,1]$. Gionis et al. study the CAMPAIGN problem, whose goal is to identify a set of target nodes $T$, whose positive opinion about an information item will maximize the overall positive opinion for the item in the social network. The objective function to maximize is therefore $g(\textbf{z})=\sum_{i=1}^n z_i$.

Bhawalkar et al. \cite{Bhawalkar:2013:COF:2488608.2488615} analyze the equilibrium outcomes of symmetric co-evolutionary game and the K-nearest neighbor (K-NN) game, distinguishing between internal and stated opinions with the usual symbols $s_i$ and $z_i$ (which are real numbers). In the K-NN game, each agent has exactly $K$ friends, so the interaction is of the nearest neighbors type and the size of $N(i)$ is exactly $K$ for each agent $i$. 

Ferraioli et al. \cite{FERRAIOLI201696} continue the study of Bindel et al. by simplifying their model to the case of binary opinion $z_i$, which can be found in the individual's voting intention in a referendum, while $s_i\in[0,1]$. They study best-response dynamics and show upper and lower bounds on the convergence to Nash equilibria.

The cost function considered by Chierichetti et al. \cite{CHIERICHETTI201811} (where update rules for $z_i$ and $s_i$ are not present) replaces the quadratic terms in \cite{BINDEL2015248} by distances in a discrete metric space while $s_i$ belongs to a discrete set (binary in some special cases)\footnote{The authors refer to this class of games as \emph{discrete preference games}.} and $z_i\in \mathbb R$. The authors adopt a strategy $\textbf{z}$ minimizing the social cost function as an optimal solution and establish bounds on the \emph{price of stability}\footnote{The price of stability is a measure of the game efficiency that is commonly adopted instead of the price of anarchy when multiple Nash equilibria are present, and is defined as on the ratio between the social cost of the best Nash equilibrium and the optimal solution. We return to the subject in Section \ref{sec:price}.}

Auletta et al. \cite{auletta2016generalized} consider a personal cost that is defined through a monotone non-decreasing function of $\textbf{z}$, assuming binary $z_i$ and $s_i$ (without update rules for them). The authors called that class \emph{generalized discrete preference games}. In particular, they show that every game with two strategies per agent that admits a generalized ordinal potential is structurally equivalent (in particular, better-response equivalent) to a generalized  discrete preference game. In another work \cite{Auletta:2017:RDP:3091282.3091307}, the same authors consider the game in which agents are utility maximizers, $z_i$, $s_i\in\{0,1\}$ and address the questions of price of stability/price of anarchy of a game in terms of the social welfare: $SW(\textbf{z}):=\sum_i u_i(\textbf{z})$.

Bil{\`o} et al. \cite{Bilo} focus on the case in which, for each player $i$, the innate opinion $s_i\in[0,1]$, while the expressed opinion $z_i\in\{0,1\}$. They define a cost-minimization $n$-player game. Bil{\`o} et al. show that any game in this class always admits an ordinal potential that implies the existence of pure Nash equilibria and convergence of better-response dynamics starting from any arbitrary strategy profile. The social optimum is obtained with respect to the problem of minimizing the sum of the players' costs. They also focus on the efficiency losses due to selfish behavior and give upper and lower bounds on the price of anarchy and lower bounds on the price of stability.

In \cite{CHEN2016808}, Chen et al. bound the price of anarchy for a game in which both $s_i$ and $z_i$ are real numbers.


\section{Agent-Based Model}
\label{sec:model}

In this Section we introduce an ABM to study the \emph{personal finance game}. In our ABM there are three classes of agents: 
\begin{itemize}
    \item a bank ($B$);
    \item a financial advisor ($A$);
    \item a set of $n$ customers or customers ($CL_i$, $i=1,\dots,n$).
\end{itemize}

Our model falls in the literature framework of \emph{opinion formation games} where game players can express some opinions and may change them according to the interactions with the other agents. For some of them the opinions they express need not coincide with their true opinions. The opinions concern investment decisions.

The aim of bank $B$ is to steer the customers towards a particular investment decision, represented by an opinion $w\in\mathbb R^+$. For example, $w$ could concern the decision to buy a security $S_1$ rather than a different security $S_2$ or other financial instruments.

The financial advisor $A$ expresses an opinion $s\in\mathbb R^+$ which need not coincide with his/her true opinion $x\in\mathbb R^+$, respectively referred to in the following as the \emph{stated opinion} and the \emph{internal opinion}. The financial advisor may therefore be \textit{untruthful}. $A$ is paid by $B$ and he/she gives advice (by way of $s$) to customers when invited to do so, but the stated opinion $s$ might not perfectly correspond to the one recommended by bank, $w$ (to preserve his/her good reputation, for instance).

customers have their opinions $c_i$,  $i=1,\dots, n$ , which fall within the range $[d_i,s]$ where $s$ is the stated opinion and $d_i\leq s$ is a positive lower bound, which represents the opinion that the customer $i$ would assume if there wasn't any interaction with A. The opinions of all customers are collected in $\textbf{c}:=(c_1,\dots,c_n)\in\mathbb R^n$.


Opinions $c_i$, $i=1,\dots, n$ and $s$ change over time, i.e. $c_i=c_i\left(t\right)$ $i=1,\dots, n$ and $s=s(t)$, while $w$, $x$ and $d$ are fixed over time. However, we assume that all the opinions lie within the range $[0,1]$.

In the spirit of the models \cite{bernheim1994theory,BUECHEL2015240}, we consider a utility function for $A$ that depends on the incentive to be truthful (the intrinsic part) and the incentive to steer the customers towards $w$ (the remunerative part). The incentive to be truthful could be related to the Advisor's conscience or to the desire of the Advisor to keep his/her reputation. Additionally, we assume that the utility function for $A$ also depends on the desire to influence the customers. The resulting utility function is supposed to be a quadratic form in the opinions and to be additive.

Thus, the utility of the financial advisor depends on the distance of his/her true opinion $x$ to his/her stated opinion $s$ as well as on the distance of the bank's desired investment decision $w$ to customers opinions $\textbf{c}$ and the distance between $s$ and $\textbf{c}$:
\begin{equation}
\label{eq:utility_A}
u_A(\textbf{c},s,w,x)=-\alpha \left(s-x\right)^2-\beta \sum_{i=1}^n\left(w-c_i\right)^2-\gamma \sum_{i=1}^n\left(s-c_i\right)^2 \, ,
\end{equation}
where $\alpha,\beta,\gamma>0$; $\beta$ is the remuneration coefficient for $A$ and is paid by bank $B$. The more customers  eventually buy the security pushed forward by the bank $B$, the more the advisor $A$ is remunerated. 

The advisor’s strategic leverage is the stated opinion $s$, and his/her aim is to maximize his/her utility: 
\begin{equation}
    \max_{s} u_A (\textbf{c},s,w,x).
\end{equation}

To define the utility of customers $CL_i$, $i=1,\dots,n$, we introduce the following returns on their investments:
\begin{itemize}
    \item $r_s$, which is the return proposed by $A$ to all the customers;
    \item $r_{d_i}$, which is the return that each customer considers that he/she can achieve through a ``good'' investment decision.
\end{itemize}
The customer $i$ would like to get $r_{d_i}$ but he/she does not completely trust herself (the customer is not assumed to be a financial expert)) and moves towards $r_s$. In general, two possible situations may occur: $r_s\leq r_{d_i}$ and $r_s > r_{d_i}$. In the first, $A$ proposes to the customer $i$ a return that is less (or equal) than expectations of $CL_i$ while in the second we have the opposite. The rationales for the two cases are respectively that the financial advisor is able to find a better investment than the customers due to his/her superior financial expertise or that customers have unrealistic expectations due to their poor knowledge of financial markets.

We assume that the utility of each customer $i$ depends on his/her lack of agreement with the advisor $A$. This \emph{cognitive dissonance} \cite{BINDEL2015248} provides customers with an incentive to modify their behavior to reduce the ``cost'' of this lack of consensus. Remember that customers have their opinions $c_i$, which fall within the range $[d_i,s]$ where advisor's stated opinion $s$ is a positive upper bound for them, and that parameter $d_i$ represents the opinion that the customer $i$ would assume if there weren't any interaction with A. Thus, the utility of CL$_i$ assumes value $r_s$ if $c_i=s$ but depends on the distance of his/her opinion $c_i$ to the advisor's stated opinion $s$ in all the other cases:
\begin{equation}
\label{eq:utility_Ci}
    u_{CL_i}(c_i,d_i,s)=r_{d_i}+\frac{c_i-d_i}{s-d_i}(r_s-r_{d_i})-\zeta (s-c_i)^2\, ,
\end{equation}
where $\zeta>0$ ($\forall i=1,\dots,n$) represents the sensitivity of customer to cognitive dissonance. Let us observe that the following consequences hold: 
\begin{enumerate}[(a)]
    \item $0\leq \frac{c_i-d_i}{s-d_i}\leq1$ since $d_{i}\le c_{i} \le s$;
    \item $r_{d_i}+\frac{c_i-d_i}{s-d_i}(r_s-r_{d_i})\geq 0$ for $c_i\in[d_i,s]$ because of consequence (a) and $r_{s},r_{d_{i}}>0$; 
    \item $\left[r_{d_i}+\frac{c_i-d_i}{s-d_i}(r_s-r_{d_i})\right]_{c_i=d_i}=r_{d_i}$ and $\left[r_{d_i}+\frac{c_i-d_i}{s-d_i}(r_s-r_{d_i})\right]_{c_i=s}=r_{s} \rightarrow u_{CL_i}(s,d_i,s)=r_s$ and $u_{CL_i}(d_i,d_i,s)= r_{d_i}-\zeta (d_i-c_i)^2$ (in the latter case the cost of lack of consensus is maximum).
\end{enumerate}

\section{Personal finance game and Nash equilibria}
\label{sec:game}

The model described in Section \ref{sec:model} describes the opinion dynamics of a financial advisor and his/her customers, where the opinions are influenced by each other's choice. This interaction can therefore be considered as a strategic game, where the players are the financial advisor and his/her customers (the bank's role is just to set the fixed aim $w$ and the incentive $\beta$) and their strategic leverages are respectively the stated opinion $s$ and the opinions $c_{i}'s$.

In this Section we solve the personal finance game by deriving the Nash equilibria. We also investigate their admissibility, i.e., their compatibility with the constraints embedded in the model.

\subsection{Nash equilibria}
\label{sec:nash}
We now find the Nash equilibrium of the $n+1$-player game ($n$ customers plus one financial advisor) using their best response functions. The best response functions aim at maximizing the players' utilities:
\begin{equation}
\label{eq:BestResp}
\begin{cases}
\underset{s}{\max}\quad u_A (\textbf{c},s,w,x)\\
\underset{c_i}{\max} \quad u_{CL_i}(c_i,d_i,s)\, , \quad i=1,\dots,n\, . 
\end{cases}
\end{equation}
For any $c_i$, we obtain the optimal $s$ by zeroing the  derivative of the utility 
\begin{align}
    \frac{\partial u_A(\textbf{c},s,w,x)}{\partial s}&=-2\alpha \left(s-x\right) -2\gamma \sum_{i=1}^n\left(s-c_i\right)\notag\\
    &=-2\alpha s+2\alpha x -2\gamma n s +2\gamma \sum_{i=1}^n c_i=0\, .
\end{align}
Turning to customers utility and fixing $s$, we obtain
\begin{equation}
        \frac{\partial u_{CL_i}(c_i,d_i,s)}{\partial c_i}=\frac{r_s-r_{d_i}}{s-d_i}+2\zeta (s-c_i)\, .
\end{equation}
Then the system (\ref{eq:BestResp}) that expresses best response functions becomes:
\begin{equation}
\label{eq:BestResp2}
\begin{cases}
-2\alpha s+2\alpha x -2\gamma n s +2\gamma \sum_{i=1}^n c_i=0 \\
\frac{r_s-r_{d_i}}{s-d_i}+2\zeta (s-c_i)=0\, .
\end{cases}
\end{equation}
The solution of the system of linear equations is the  advisor's best response function 
\begin{equation}
    s=\frac{1}{\alpha  +\gamma n} \left(\alpha x  +\gamma \sum_{i=1}^n c_i\right)\, ,
\end{equation}
which is a linear function of the customers' opinions.
Similarly, the customers' best response function is given by
\begin{equation}
c_i=\frac{r_s-r_{d_i}}{2\zeta (s-d_i)}+s \, .
\end{equation}
For the sake of simplicity, consider the \emph{special case} where all the customers have the same initial opinion and expectations, i.e., $r_{d_i}=r_{d}$ and $d_i=d$ $\forall i=1,\dots,n$, so that all the customers take the same investment decision, i.e., $c_i=c$. Sometimes we will denote this case as the case of \emph{homogeneous investors}.

\begin{figure}
    \centering
\begin{tikzpicture}
 \begin{axis}[xmin=0,xmax=1,ymin=0.11,ymax=1, samples=1000, xlabel={Customer's opinion},
 ylabel={Advisor's stated opinion},unbounded coords=discard,legend pos=south east]
 \addlegendimage{empty legend}
 \addlegendentry{}
  \addplot+[blue,samples=150] (x,{(0.05*0.4+0.2*1*x)/(0.05+0.2*1)});
  \addlegendentry{financial advisor}
  \addplot+[red,domain=0.11:1, samples=25] ({-0.1/(2*30*(x - 0.1))+x},x);
   \addlegendentry{customer}
 \end{axis}
\end{tikzpicture}
    \caption{Best response functions of financial advisor (blue) and customers (red). Parameters: $r_s-r_{d}=-0.1$, $\zeta=10$, $d=0.1$, $\alpha=0.05$, $x=0.4$, $\gamma=0.2$, $n=1$.}
    \label{fig:best_resp}
\end{figure}
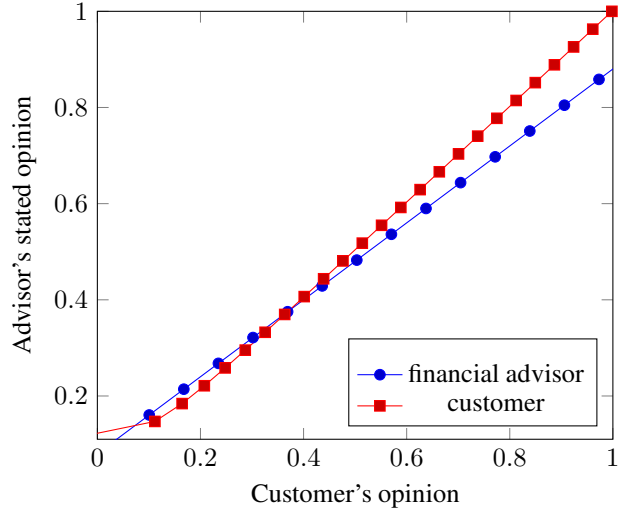

Then, the best response functions become simply
\begin{equation}
\label{eq:best_s}
    s=\frac{\alpha x  +\gamma n c}{\alpha  +\gamma n}
\end{equation}
and
\begin{equation}
\label{eq:best_c}
c=\frac{r_s-r_{d}}{2\zeta (s-d)}+s \, .
\end{equation}
We plot the best response functions of both players in Fig. \ref{fig:best_resp} for a sample case. We see that the best response function of the financial advisor in Equation (\ref{eq:best_s}) is a linear function of the customers' opinion, with a slope $\frac{\gamma n}{\alpha  +\gamma n}<1$, while the best response function of the customers in Equation (\ref{eq:best_c}) is the sum of an angle bisector and a homographic function with a vertical asymptote at $s=d$. The Nash equilibria are represented by the intersection of the two curves. Note the presence of two Nash equilibria.

In order to compute the the equilibria, we solve the system (\ref{eq:BestResp2}) for $d_i=d$. Here, by substitution  we obtain
\begin{equation}
    -2\alpha s+2\alpha x +\frac{\gamma}{\zeta} n \frac{r_s-r_{d}}{s-d}=0\, ,
\end{equation}
from which
\begin{equation}
\label{eq:s-second-deg}
2\alpha  s^2-2\alpha\left(  d+  x \right) s+ 2\alpha x d -\frac{\gamma n}{\zeta} (r_s-r_{d})=0\, ,
\end{equation}
for $s\neq d$. Accordingly, the two Nash equilibria are:
\begin{align}
\label{eq:nash}
P^\ast=(s^\ast,c_1^\ast,\dots,c_n^\ast)&=\left(a, \frac{1}{2\zeta}\frac{r_s-r_{d}}{a -d}+ a, \dots, \frac{1}{2\zeta}\frac{r_s-r_{d}}{a -d}+ a \right) \notag \\
P^\dagger=(s^\dagger,c_1^\dagger,\dots,c_n^\dagger)&= \left(b, \frac{1}{2\zeta}\frac{r_s-r_{d}}{b -d}+ b, \dots, \frac{1}{2\zeta}\frac{r_s-r_{d}}{b -d}+ b \right)\, ,
\end{align}
where
\begin{align}
\label{eq:a-b}
    a&=\frac{d+  x}{2} + \frac{1}{2} \sqrt{\left(  d-  x \right)^2  +\frac{2\gamma n}{\alpha\zeta} (r_s-r_{d}) }\notag \\
    b&=\frac{d+  x}{2} - \frac{1}{2} \sqrt{\left(  d-  x \right)^2  +\frac{2\gamma n}{\alpha\zeta} (r_s-r_{d}) }
\end{align}
are the roots of quadratic equation (\ref{eq:s-second-deg}).

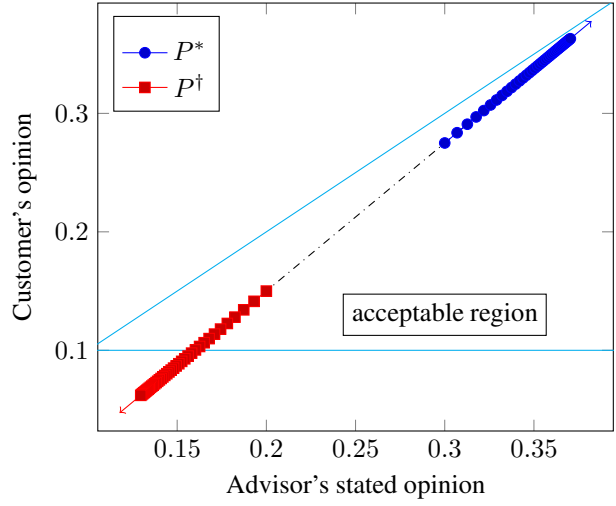
\begin{figure}
\centering
\begin{tikzpicture}
\begin{axis}[xlabel={Advisor's stated opinion},ylabel={Customer's opinion},legend pos=north west]
\draw[cyan] (0.1,0.1) 
  -- (1,0.1) 
  -- (1,1)
  -- cycle;
\draw[dash dot] (0.370416, 0.36302) -- (0.129584, 0.0619801);
\node[draw] at (0.3,0.13) {acceptable region};
\draw[->,red] (0.129584,0.0619801) -- (0.117712,0.0471405); 
\draw[->,blue] (0.370416,0.36302) -- (0.382288,0.377859); 
\addlegendimage{empty legend}
\addlegendentry{}
\addplot+[blue, samples=40, variable=\zeta, domain=10:25]
({(0.1 + 0.4)/2 + 1/2*sqrt((0.1 - 0.4)^2 + 2*0.2*1/(0.05*zeta)*(-0.1))},
{1/(2*zeta)*(-0.1)/((0.1 + 0.4)/2 + 1/2*sqrt((0.1 - 0.4)^2 + 
2*0.2*1/(0.05*zeta)*(-0.1)) - 0.1) + (0.1 + 0.4)/2 +  1/2*sqrt((0.1 - 0.4)^2 
+ 2*0.2*1/(0.05*zeta)*(-0.1))});
\addlegendentry{$P^\ast$}
\addplot+[red, samples=40, variable=\zeta, domain=10:25]
({(0.1 + 0.4)/2 - 1/2*sqrt((0.1 - 0.4)^2 + 2*0.2*1/(0.05*zeta)*(-0.1))},
{1/(2*zeta)*(-0.1)/((0.1 + 0.4)/2 - 1/2*sqrt((0.1 - 0.4)^2 + 
2*0.2*1/(0.05*zeta)*(-0.1)) - 0.1) + (0.1 + 0.4)/2 -  1/2*sqrt((0.1 - 0.4)^2 
+ 2*0.2*1/(0.05*zeta)*(-0.1))});
\addlegendentry{$P^\dagger$}
\end{axis}
\end{tikzpicture}   
\caption{Nash equilibria' dependence on increasing values of $\zeta$. Fixed parameters: $r_s-r_{d}=-0.1$, $d=0.1$, $\alpha=0.05$, $x=0.4$, $\gamma=0.2$, $n=1$. For $\zeta=5$ we obtain no real Nash equilibria.}
    \label{fig:nash_equilibria_parameters}
\end{figure}

\begin{figure}
    \centering
\begin{tikzpicture}
\begin{axis}[xlabel={Advisor's stated opinion},ylabel={Customer's opinion},legend pos=south east]
\draw[cyan] (0.1,0.1) 
  -- (1,0.1) 
  -- (1,1)
  -- cycle;
\node[draw] at (0.32,0.13) {acceptable region};
\draw[->,red] (0.103371,-1.37977) -- (0.102804,-1.48037); 
\draw[->,blue] (0.388444,0.371109) -- (0.399666,0.382981); 
\addlegendimage{empty legend}
\addlegendentry{}
\addplot+[blue, samples=40, variable=\alpha, domain=0.05:0.25]
    ({(0.1 + 0.4)/2 + 1/2*sqrt((0.1 - 0.4)^2 + 2*0.2*1/(alpha*10)*(-0.1))},
    {1/(2*10)*(-0.1)/((0.1 + 0.4)/2 + 1/2*sqrt((0.1 - 0.4)^2 + 2*0.2*1/(alpha*10)*(-0.1)) - 0.1) + (0.1 + 0.4)/2 +  1/2*sqrt((0.1 - 0.4)^2 + 2*0.2*1/(alpha*10)*(-0.1))});
    \addlegendentry{$P^\ast$}
\addplot+[red, samples=40, variable=\alpha, domain=0.05:1]
    ({(0.1 + 0.4)/2 - 1/2*sqrt((0.1 - 0.4)^2 + 2*0.2*1/(alpha*10)*(-0.1))},
    {1/(2*10)*(-0.1)/((0.1 + 0.4)/2 - 1/2*sqrt((0.1 - 0.4)^2 + 2*0.2*1/(alpha*10)*(-0.1)) - 0.1) + (0.1 + 0.4)/2 -  1/2*sqrt((0.1 - 0.4)^2 + 2*0.2*1/(alpha*10)*(-0.1))});
    \addlegendentry{$P^\dagger$}
\end{axis}
\end{tikzpicture}
    \caption{Nash equilibria' dependence on increasing values of $\alpha$. Fixed parameters: $r_s-r_{d}=-0.1$, $d=0.1$, $\zeta=10$, $x=0.4$, $\gamma=0.2$, $n=1$.}
    \label{fig:nash_equilibria_parameters2}
\end{figure}
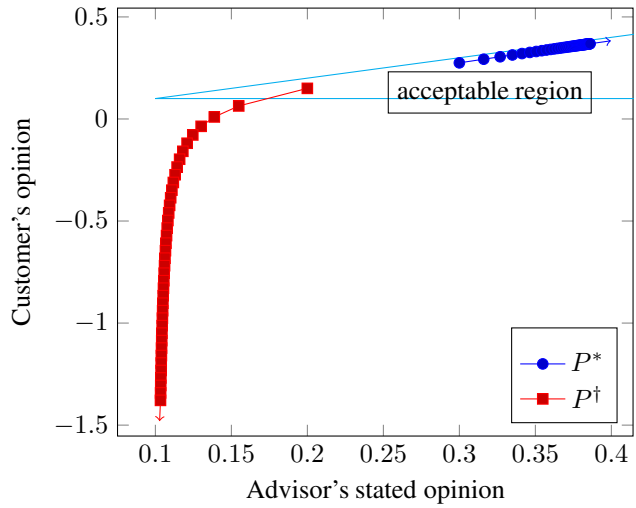

\begin{figure}
    \centering
\begin{tikzpicture}
\begin{axis}[xlabel={Advisor's stated opinion},ylabel={Customer's opinion},legend pos=south east]
\draw[cyan] (0.1,0.1) 
  -- (1,0.1) 
  -- (1,1)
  -- cycle;
\node[draw] at (0.305,0.15) {acceptable region};
\draw[->,red]  (0.235858,0.199055) -- (0.25,0.216667);
\draw[->,blue] (0.3,0.275) -- (0.255,0.223);
\addlegendimage{empty legend}
\addlegendentry{}
\addplot+[blue, samples=40, variable=\gamma, domain=0.05:0.223]
    ({(0.1 + 0.4)/2 + 1/2*sqrt((0.1 - 0.4)^2 + 2*gamma*1/(0.05*10)*(-0.1))},
    {1/(2*10)*(-0.1)/((0.1 + 0.4)/2 + 1/2*sqrt((0.1 - 0.4)^2 + 2*gamma*1/(0.05*10)*(-0.1)) - 0.1) + (0.1 + 0.4)/2 +  1/2*sqrt((0.1 - 0.4)^2 + 2*gamma*1/(0.05*10)*(-0.1))});
    \addlegendentry{$P^\ast$}
\addplot+[red, samples=40, variable=\gamma, domain=0.05:0.223]
    ({(0.1 + 0.4)/2 - 1/2*sqrt((0.1 - 0.4)^2 + 2*gamma*1/(0.05*10)*(-0.1))},
    {1/(2*10)*(-0.1)/((0.1 + 0.4)/2 - 1/2*sqrt((0.1 - 0.4)^2 + 2*gamma*1/(0.05*10)*(-0.1)) - 0.1) + (0.1 + 0.4)/2 -  1/2*sqrt((0.1 - 0.4)^2 + 2*gamma*1/(0.05*10)*(-0.1))});
    \addlegendentry{$P^\dagger$}
\end{axis}
\end{tikzpicture}
    \caption{Nash equilibria' dependence on decreasing values of $\gamma$. Fixed parameters: $r_s-r_{d}=-0.1$, $d=0.1$, $\zeta=10$, $x=0.4$, $\alpha=0.05$, $n=1$. For $\gamma=0.3$ we obtain no real Nash equilibria.}
    \label{fig:nash_equilibria_parameters3}
\end{figure}
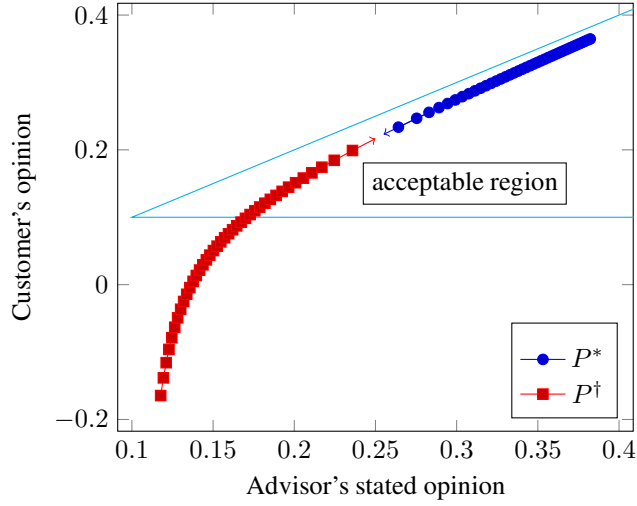

We can now examine the dependence of the Nash equilibria on the model parameters, recalling that $\gamma$ measures the importance of the advisor's influence on customers, $\alpha$ measures the importance of truthfulness, $\beta$ measures the importance of remuneration for the advisor's choice, and $\zeta$ measures the importance of belief in the advisor's stated opinion (i.e., the cognitive dissonance). The curves are shown in Figs. \ref{fig:nash_equilibria_parameters}-\ref{fig:nash_equilibria_parameters3}, where the parameters are held fixed excepting that of interest. The  curves shows how the two equilibria move, with an arrow indicating the direction of growth of the parameter of interest. A triangular region is shown as bounded by the two straight lines: equilibria falling outside that region are not acceptable since they violate the constraint on customers' opinion ($d\le c \le s$).   

\textit{Impact of cognitive dissonance}\\
In response to changes in $\zeta$, Nash equilibria are placed along the dash-dotted line in Fig. \ref{fig:nash_equilibria_parameters}. When $\zeta$ increases, both equilibria tend to pull away and to accumulate in different areas. This is because, if we fix $n=1$ and assume e.g. $d<x$, as $\zeta$ tends to $\infty$ the equilibria $P^\ast$ and $P^\dagger$ in (\ref{eq:nash}) become
    \begin{align}
    \label{eq:dissonance}
        &\lim_{\zeta \to +\infty}\left(a, \frac{1}{2\zeta}\frac{r_s-r_{d}}{a -d}+ a\right)=\left(x,  x\right)\notag \\
        &\lim_{\zeta \to +\infty}\left(b, \frac{1}{2\zeta}\frac{r_s-r_{d}}{b -d}+ b\right)=\left(d, d- \frac{\alpha(x-d)}{\gamma}\right)\, .
    \end{align}
As $\zeta$ tends to $\infty$, the second of (\ref{eq:dissonance}) represents an unreachable limit because there exists a number $\bar\zeta>0$ such that, for each $\zeta>\bar\zeta$, the support of the curve
\begin{equation}
    \left(b, \frac{1}{2\zeta}\frac{r_s-r_{d}}{b -d}+ b\right) \quad \text{where} \quad b=\frac{d+  x}{2} - \frac{1}{2} \sqrt{\left(  d-  x \right)^2  +\frac{2\gamma n}{\alpha\zeta} (r_s-r_{d}) }
\end{equation}
goes out the triangular acceptance region (see Fig. \ref{fig:nash_equilibria_parameters}, the squared branch). If we assume that, for $\zeta=\bar\zeta$, the curve $ \left(b, \frac{1}{2\zeta}\frac{r_s-r_{d}}{b -d}+ b\right)$ falls right onto the horizontal side of the triangle, we define
\begin{equation}
    \left(b, \frac{1}{2\bar\zeta}\frac{r_s-r_{d}}{b -d}+ b\right) \quad \text{where} \quad b=\frac{d+  x}{2} - \frac{1}{2} \sqrt{\left(  d-  x \right)^2  +\frac{2\gamma n}{\alpha\bar\zeta} (r_s-r_{d}) }
\end{equation}
the \emph{last useful equilibrium}. The \emph{critical value} $\bar\zeta$ can be found by imposing
\begin{equation}
\label{eq:lastequi}
    \frac{1}{2\bar\zeta}\frac{r_s-r_{d}}{b -d}+ b=d \quad \text{where} \quad b=\frac{d+  x}{2} - \frac{1}{2} \sqrt{\left(  d-  x \right)^2  +\frac{2\gamma n}{\alpha\bar\zeta} (r_s-r_{d}) }
\end{equation}
because $c=d$ represents the horizontal side of the triangular acceptance region (let us remember that $d\le c \le s$). Then, from (\ref{eq:lastequi}) it is easily to prove that
\begin{equation}
\label{eq:criticalvalue}
    \bar\zeta=-\frac{1}{2}\left(\frac{\alpha+\gamma}{\alpha}\right)^2 \frac{r_s-r_d}{(x-d)^2}.
\end{equation}
The value in (\ref{eq:criticalvalue}) is positive if $r_s<r_d$. Calculated in (\ref{eq:criticalvalue}), the last useful equilibrium is
\begin{equation}
    \left(\frac{d+x}{2}-\frac{1}{2}\frac{\left|(d-x)(\alpha-\gamma)\right|}{\alpha+\gamma}, d\right)\, .
\end{equation}
\begin{remark}
For the sake of simplicity, Eq. (\ref{eq:dissonance}) have been obtained for the particular case $d<x$. Anyway, it is possible to prove that
    \begin{align}
    \label{eq:dissonance2}
        \lim_{\zeta \to +\infty}&\left(a, \frac{1}{2\zeta}\frac{r_s-r_{d}}{a -d}+ a\right)=\left(\frac{d+x}{2}+\frac{|d-x|}{2},\right.\notag \\  
        &\left. \frac{d+x}{2}+\frac{|d-x|}{2}-\alpha \frac{x-d-|x-d|}{2\gamma}\right)\notag \\
        \lim_{\zeta \to +\infty}&\left(b, \frac{1}{2\zeta}\frac{r_s-r_{d}}{b -d}+ b\right)= \left(\frac{d+x}{2}-\frac{|d-x|}{2},\right.\notag \\ &\left.\frac{d+x}{2}-\frac{|d-x|}{2}-\alpha \frac{x-d+|x-d|}{2\gamma}\right)\, .
    \end{align}

\end{remark}
    
\textit{Impact of advisor's truthfulness}\\
A similar result also applies to the case $\alpha\to +\infty$. Actually, if we assume $d<x$, as $\alpha$ tends to $\infty$ the equilibria $P^\ast$ in (\ref{eq:nash}) is
    \begin{equation}
        \lim_{\alpha \to +\infty}\left(a, \frac{1}{2\zeta}\frac{r_s-r_{d}}{a -d}+ a\right)=\left(x,  x\right)    
    \end{equation}
    while $P^\dagger$ is excluded because this time
    \begin{equation}
        \lim_{\alpha \to +\infty} \frac{1}{2\zeta}\frac{r_s-r_{d}}{b -d}+ b=\infty\, .
    \end{equation}
    This case corresponds to an advisor that is very sensitive to the difference between his/her stated and true opinion, i.e. to the difference between what he/she says and what he/she really thinks. See Fig. \ref{fig:nash_equilibria_parameters2}.

\textit{Impact of trust in the advisor}\\ 
Fig. \ref{fig:nash_equilibria_parameters3} shows that, as $\gamma$ approaches to zero, only one equilibria survives. Actually, for $n=1$ and $d<x$, as $\gamma\to0$ the equilibria $P^\ast$ in (\ref{eq:nash}) is
    \begin{equation}
        \lim_{\gamma \to 0}\left(a, \frac{1}{2\zeta}\frac{r_s-r_{d}}{a -d}+ a\right)=\left(x, \frac{r_s-r_d}{2\zeta(x-d)}+ x\right)    
    \end{equation}
    while $P^\dagger$ is excluded because
    \begin{equation}
        \lim_{\gamma \to 0} \frac{1}{2\zeta}\frac{r_s-r_{d}}{b -d}+ b=\infty\, .
    \end{equation}
    Since $\gamma$ tells us how wide is the advisor's desire to influence the customers, when $\gamma=0$ (and all the other parameters are fixed and $\neq 0$) his/her equilibrium is represented by his/her internal opinion. (There is no desire to influence the customers, then there is no reason to tell a lie.) Observe also that $\frac{r_s-r_d}{2\zeta(x-d)}+ x<x$ if $r_s<r_d$; in this situation the Nash equilibrium is acceptable and the equilibrium solution for customer is different from the advisor's internal opinion.

\begin{remark}
In the above bullet list we have considered $\alpha$, $\zeta\to + \infty$ and $\gamma\to 0$. The same considerations remain true, in approximation, if we substitute $\alpha$, $\zeta\to + \infty$ and $\gamma\to 0$ with finite values such that $\frac{2\gamma n}{\alpha\zeta} \, \frac{r_s-r_{d}}{\left(  d-  x \right)^2}\ll 1$. E.g. in (\ref{eq:a-b}):
\begin{equation}
    \sqrt{1  +\frac{2\gamma n}{\alpha\zeta} \, \frac{r_s-r_{d}}{\left(  d-  x \right)^2} } \approx 1+\frac{\gamma n}{\alpha\zeta} \, \frac{r_s-r_{d}}{\left(  d-  x \right)^2}+\ldots
\end{equation}
whenever $\frac{2\gamma n}{\alpha\zeta} \, \frac{|r_s-r_{d}|}{\left(  d-  x \right)^2}<1$.
\end{remark}

\begin{figure}
    \centering
\begin{tikzpicture}
\begin{axis}[xlabel={$s$},ylabel={$c$},legend pos=south east]
\draw[cyan] (0.1,0.1) 
  -- (1,0.1) 
  -- (1,1)
  -- cycle;
\node[draw] at (0.62,0.55) {acceptable region};
\draw[->,red]  (0.538763,0.525864) -- (0.55,0.54);
\draw[->,blue] (0.661237,0.658136) -- (0.65,0.646667);
\addlegendimage{empty legend}
\addlegendentry{}
\addplot+[blue, samples=40, variable=\gamma, domain=10:100]
    ({(0.5 + 0.7)/2 + 1/2*sqrt((0.5 - 0.7)^2 + 2*gamma*1/(8*100)*(-0.1))},
    {1/(2*100)*(-0.1)/((0.5 + 0.7)/2 + 1/2*sqrt((0.5 - 0.7)^2 + 2*gamma*1/(8*100)*(-0.1)) - 0.5) + (0.5 + 0.7)/2 +  1/2*sqrt((0.5 - 0.7)^2 + 2*gamma*1/(8*100)*(-0.1))});
    \addlegendentry{$P^\ast$}
\addplot+[red, samples=40, variable=\gamma, domain=10:100]
    ({(0.5 + 0.7)/2 - 1/2*sqrt((0.5 - 0.7)^2 + 2*gamma*1/(8*100)*(-0.1))},
    {1/(2*100)*(-0.1)/((0.5 + 0.7)/2 - 1/2*sqrt((0.5 - 0.7)^2 + 2*gamma*1/(8*100)*(-0.1)) - 0.5) + (0.5 + 0.7)/2 -  1/2*sqrt((0.5 - 0.7)^2 + 2*gamma*1/(8*100)*(-0.1))});
    \addlegendentry{$P^\dagger$}
\end{axis}
\end{tikzpicture}
    \caption{Nash equilibria' dependence on $\gamma$, like in Fig. \ref{fig:nash_equilibria_parameters3} but for increasing values of the parameter. Fixed parameters: $r_s-r_{d}=-0.1$, $d=0.5$, $\zeta=100$, $x=0.7$, $\alpha=8$, $n=1$.}
    \label{fig:nash_equilibria_parameters3bis}
\end{figure}
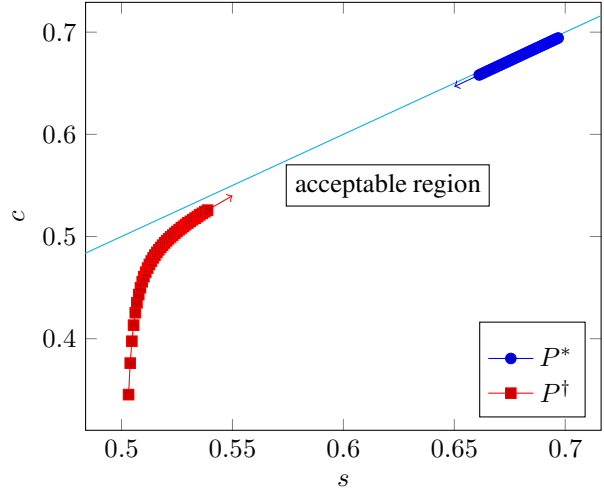

Let us conclude this section by considering the dependence of Nash equilibria on the increasing measure of importance of the influence on customers, $\gamma$, which is depicted in Fig. \ref{fig:nash_equilibria_parameters3bis}. Observe that, if $\gamma$ gets too big, both equilibria become not real.

\subsubsection{Admissibility of the Nash equilibria}
\label{sec:adm_nash}

In this Section we derive the parameter region in which Nash equilibria, obtained in Section \ref{sec:nash}, stay coherent with opinion variable definition. For example, since $c_1$, $\dots$, $c_n$, $s$ shall fall within the range $[0,1]$, the coordinates of Nash equilibria are constrained between $0$ and $1$, and hence we obtain the conditions on parameters to ensure that. We will focus on the \emph{special case}, i.e. $r_{d_i}=r_{d}$ and $d_i=d$ $\forall i=1,\dots,n$.

In our model we distinguish between strategic variables and parameters. The first of these are $c_i$, $i=1,2,\dots,n$, and $s$, while the latter are $d$, $x$, $w$, $n$, $\alpha$, $\beta$, $\gamma$, $\zeta$, $r_d$, $r_s$. The set of all numeric values that they can assume are called, respectively, the \emph{domain} $\mathcal{D}$ and the \emph{admissible parameter region} $\mathcal{R}$.

From the assumptions on opinion variables and parameters, we have that
\begin{equation}
\label{eq:domain1}
    \mathcal{D}:=\left\{(c_1,\dots,c_n,s)\, \big|\, d \leq c_i\leq s\, \forall i=1,\dots,n,\, s\in[d,1]\right\}\, ,
\end{equation}
\begin{align}
\label{eq:a_region}
    \mathcal{R}:=&\left\{(d,x,w,n,\beta,\gamma,\zeta,\alpha,r_s,r_d)\, \big|\, d,x,w\in[0,1]\, , \right.\notag \\ 
    &\left. n\in\mathbb N\, , \, \beta,\gamma,\zeta,\alpha>0\, , \, r_d,r_s\in[0,1]\right\}\, .
\end{align}
Let us denote with
\begin{equation}
    Y^{n}=\underbrace {Y\times Y\times \cdots \times Y} _{n}=\{(y_{1},\ldots ,y_{n})\ |\ y_{i}\in Y\ {\text{for every}}\ i\in \{1,\ldots ,n\}\}
\end{equation}
the $n$-ary Cartesian power of a set $Y$. Hence, the domain can be rewritten as
\begin{equation}
\label{eq:domain_cart}
    \mathcal{D}=[d,s]^n\times [d,1]\, , 
\end{equation}

Let us derive conditions on the parameters (in other words, subsets of $\mathcal{R}$) which ensure the existence of $P^\ast$ and $P^\dagger$, by distinguishing between two cases, \textbf{(A)} $r_s=r_d$ and \textbf{(B)} $r_s\neq r_d$.

\textbf{Case (A)}, $r_s=r_d$. In view of this, Nash equilibria becomes:
\begin{align}
\label{eq:nashA}
P^\ast=(s^\ast,c_1^\ast,\dots,c_n^\ast)&=\left(\frac{d+x}{2}+\frac{|d-x|}{2},  \dots, \frac{d+x}{2}+\frac{|d-x|}{2} \right) \notag \\
P^\dagger=(s^\dagger,c_1^\dagger,\dots,c_n^\dagger)&= \left(\frac{d+x}{2}-\frac{|d-x|}{2},  \dots,  \frac{d+x}{2}-\frac{|d-x|}{2} \right)\, .
\end{align} 
We have the following result:
\begin{proposition}
Let $r_s=r_d$. The \emph{Personal Finance Game} admits the following Nash equilibria:
\begin{equation}
    \left\{\begin{array}{llcr}
        P^\ast=\left(x,  \dots, x\right), & P^\dagger=\left(d,  \dots,  d \right) & \text{for} & d<x\\
        & P^\ast=P^\dagger & \text{for} & d=x\\
        & P^\ast=\left(d,  \dots,  d \right) & \text{for} & d>x
        \end{array}\right.
\end{equation}
\end{proposition}
\proof
By virtue of constraints (\ref{eq:domain1}), the admissible parameter regions in which $P^\ast$ and $P^\dagger$ are acceptable Nash equilibria are described by:
\begin{equation}
\begin{array}{cc}
       \mathcal{R}_1^\ast=\left\{  0\leq d \leq \frac{d+x}{2}+\frac{|d-x|}{2}\leq 1\right\}\, , &      \mathcal{R}_1^\dagger=\left\{0\leq d \leq \frac{d+x}{2}-\frac{|d-x|}{2}\leq 1\right\}
\end{array}
\end{equation}
and

where we denoted by $\mathcal{R}_1^\ast$, $\mathcal{R}_1^\dagger \subseteq \mathcal{R}$ these regions ($\mathcal{R}_1^\ast$ for $P^\ast$ and $\mathcal{R}_1^\dagger$ for $P^\dagger$). $\square$

\textbf{Case (B)}, $r_s\neq r_d$. Let us denoted by $\mathcal{R}_1^\ast$, $\mathcal{R}_1^\dagger \subseteq \mathcal{R}$ the admissible parameter regions in which, respectively, $P^\ast$ and $P^\dagger$ are acceptable Nash equilibria. Then
\begin{proposition}
\label{p:pv}
Let $r_s\neq r_d$, then
\begin{align}
\label{eq:adm_ast}
\mathcal{R}_1^\ast=&[0,x)\times[0,1]^2\times \mathbb N\times (0,+\infty)^3\times&\notag\\ 
&\times\left[ \left( (0,\gamma n)\times [r_d-r_d^{(1)},r_d) \times [0,1] \right) \right. \notag \\
&\left. \cup \left( [\gamma n,+\infty)\times [r_d-r_d^{(2)},r_d) \right)\times [0,1] \right]
\end{align}
and
\begin{equation}
\label{eq:adm_ast-copy}
\mathcal{R}_1^\dagger=[0,x)\times[0,1]^2\times \mathbb N\times (0,+\infty)^3\times  (0,\gamma n)\times [r_d-r_d^{(1)},r_d-r_d^{(2)}] \times [0,1] 
\end{equation}
where with $r_d^{(1)},r_d^{(2)}$ we denoted, respectively, $\frac{\alpha\zeta}{2\gamma n}(x-d)^2$ and $2\zeta \alpha^2 \left(\frac{x-d}{\alpha+\gamma n}\right)^2$. 
\end{proposition}
\proof
The completed proof is given in Appendix. $\square$

\begin{remark}
\label{r:2}
The intersection $\mathcal{R}_1^\dagger\cap \mathcal{R}_1^\ast$ is not empty. Then, for parameters values in $\mathcal{R}_1^\dagger\cap \mathcal{R}_1^\ast$, the two Nash equilibria $P^\ast$ and $P^\dagger$ are both acceptable (see e.g. Fig. \ref{fig:best_resp}).
\end{remark}

\begin{remark}
\label{r:1}
Focusing on $r_s$ range in the equation of admissible parameter region $\mathcal{R}_1^\ast\cup \mathcal{R}_1^\dagger$, the equations (\ref{eq:adm_ast}) and (\ref{eq:adm_ast-copy}), we notice that $r_s< r_d$. The customer $i$, that in our paper is a small investor, could be naive about incentives and expects a return bigger than the one advisor A proposes instead to her. However, A, who can choose to express an opinion which need not coincide with his/her true opinion, is supposed to be \emph{unbiased}: they do not manipulate the expectations of naive investors, which can be translated in the condition $r_s\neq r_d$. See, e.g., \cite{HONG2008268} for the role of advisors and their communication process with investors in generating divergence of opinion, \cite{MALMENDIER2007457} for evidence on investor reaction to recommendations and \cite{Hong03} for evidence on analyst incentives.
\end{remark}

\subsection{The boundary of $\mathcal{D}$}
\label{sec:boundary}

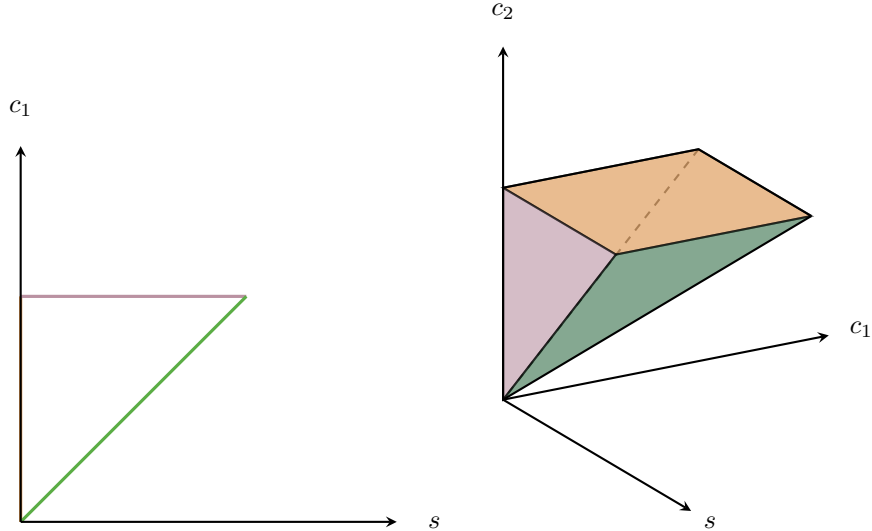
\begin{figure}[ht]
\begin{center}
\begin{tikzpicture}[thick,scale=5]
\coordinate (A1) at (0,0);
\coordinate (A2) at (0,0.6);
\coordinate (A3) at (0.6,0.6);
\draw[color=cof,very thick] (A1) -- (A2);
\draw[color=pur,very thick] (A2) -- (A3);
\draw[color=greeo,very thick] (A3) -- (A1);
\draw[-stealth] (0,0) -- (1,0) node[pos=1.1]{$s$};
\draw[-stealth] (0,0) -- (0,1) node[pos=1.1]{$c_1$};
\end{tikzpicture}
\quad
\begin{tikzpicture}[thick,scale=5]
\tdplotsetmaincoords{70}{60}
\begin{scope}[tdplot_main_coords]
\coordinate (A1) at (0,0,0.6);
\coordinate (A2) at (0,0.6,0.6);
\coordinate (A3) at (0.6,0.6,0.6);
\coordinate (A4) at (0.6,0,0.6);
\coordinate (B2) at (0,0,0);

\begin{scope}[thick,dashed,,opacity=0.6]
\draw (A2) -- (B2);
\end{scope}
\draw[fill=cof,opacity=0.6] (A1) -- (A4) -- (A3) -- (A2) -- cycle;
\draw[fill=pur,opacity=0.6] (A1) -- (A4) -- (B2);
\draw[fill=greet,opacity=0.6] (A3) -- (A4) -- (B2);
\draw (A1) -- (B2) -- (A3);
\draw (A1) -- (A2) -- (A3);
\draw[-stealth] (0,0,0) -- (1,0,0) node[pos=1.1]{$s$};
\draw[-stealth] (0,0,0) -- (0,1,0) node[pos=1.1]{$c_1$};
\draw[-stealth] (0,0,0) -- (0,0,1) node[pos=1.1]{$c_2$};
\end{scope}
\end{tikzpicture}
\end{center}
\caption{Graphical representation of $\partial \mathcal{D}$ for two ($n=1$, on the left) and three ($n=2$, on the right) dimensional spaces. The boundaries are highlighted in different colors.}
\label{fig:3}
\end{figure}

The following result characterizes mathematically the boundary of the set $\mathcal{D}$ described in (\ref{eq:domain1}) and denoted by $\partial \mathcal{D}$. A graphical representation of $\partial \mathcal{D}$ for two and three dimensional spaces is depicted in Fig. \ref{fig:3}, where the boundaries are highlighted in different colors. Mathematically, since 
\begin{equation}
    \mathcal{D}:=\left\{(c_1,s)\, \big|\, d \leq c_1\leq s\leq 1\right\}\, , \quad n=1
\end{equation}
and
\begin{equation}
    \mathcal{D}:=\left\{(c_1,c_2,s)\, \big|\, d \leq c_1\leq s\leq 1\, , \, d \leq c_2\leq s\leq 1\right\}\, , \quad n=2
\end{equation}
they are described respectively by
\begin{equation}
\label{eq:partiald1}
    \partial\mathcal{D}:=\left\{c_1=d\, , \, s\in[d,1]\right\}\cup \left\{s=1\, , \, c_1\in[d,1]\right\}\cup \left\{s = c_1\, , \, c_1\in[d,1]\right\}
\end{equation}
for $n=1$, and
\begin{equation}
    \partial\mathcal{D}:=\left\{c_1=d\, , \, s\in[d, 1]\, , \, d \leq c_2\leq s\right\}\cup \left\{d \leq c_1\leq s\, , \, c_2=d\, , \, s\in[d,1]\right\}
\end{equation}
\begin{equation}
\cup \left\{s= 1\, , \, c_1,c_2\in[d,1]\right\}\cup \left\{s=c_1\, , \, c_1\in[d,1] \, , \, d \leq c_2\leq c_1\right\}
\end{equation}
\begin{equation}
\cup \left\{s=c_2\, , \, c_2\in[d,1] \, , \, d \leq c_1\leq c_2\right\}
\end{equation}
for $n=2$. Another example (four dimensional space) has equation
\begin{equation}
    \mathcal{D}=\left\{(c_1,c_2,c_3,s)\, \big|\, d \leq c_1\leq s\leq 1\, , \, d \leq c_2\leq s\leq 1\, , \, d \leq c_3\leq s\leq 1 \right\}\, ,
\end{equation}
for $n=3$ and then
\begin{equation}
    \partial\mathcal{D}:=\left\{c_1=d\, , \, s\in[d, 1]\, , \, d \leq c_2\leq s\, , \, d \leq c_3\leq s\right\}\cup 
\end{equation}
\begin{equation}
\cup    \left\{c_2=d\, , \, s\in[d, 1]\, , \, d \leq c_1\leq s\, , \, d \leq c_3\leq s\right\}\cup
\end{equation}
\begin{equation}
\cup    \left\{c_3=d\, , \, s\in[d, 1]\, , \, d \leq c_1\leq s\, , \, d \leq c_2\leq s\right\}\cup \left\{s= 1\, , \, c_1,c_2,c_3\in[d,1]\right\}\cup
\end{equation}
\begin{equation}
\cup  \left\{s=c_1\, , \, c_1\in[d,1] \, , \, d \leq c_2\leq c_1\, , \, d \leq c_3\leq c_1\right\}
\end{equation}
\begin{equation}
\cup \left\{s=c_2\, , \, c_2\in[d,1] \, , \, d \leq c_1\leq c_2\, , \, d \leq c_3\leq c_2\right\}
\end{equation}
\begin{equation}
\cup \left\{s=c_3\, , \, c_3\in[d,1] \, , \, d \leq c_1\leq c_3\, , \, d \leq c_2\leq c_3\right\}\, .
\end{equation}
\begin{proposition}
\label{p:2}
The boundary of domain $\mathcal{D}$, described by (\ref{eq:domain1}), is
\begin{align}
\label{eq:lemma_1}
   \partial\mathcal{D}= \Bigl\{(c_1,&\dots,c_n,s)\in[d,1]^{n+1} : \notag \\
   &\max_{i=1,\dots,n} c_i \leq s \;\wedge\; \left[\prod_{i=1}^n (c_i-d)\, (s-c_i)\right]\, (s-1)=0 \Bigr\}\, .
\end{align}
\end{proposition}
\proof
Let us consider the set (\ref{eq:domain1}), where $d \leq c_i\leq s\leq 1$ $\forall i=1,\dots,n$. Fix $j\in\{1,2,\dots,n\}$ and assume $\max_{i=1,\dots,n} c_i=c_j$. 

Fix, for example, $c_1=d$ then we still have $d \leq c_i\leq s\leq 1$, i.e. $d \leq c_i\leq s$ $\wedge$ $d \leq s\leq 1$, $\forall i=2,\dots,n$. Relation $d \leq s\leq 1$ and $d \leq c_i\leq 1$ are verified by definition while the truthfulness of $c_i\leq s$ is ensured by $c_j \leq s$, because $c_i\leq c_j$ $\forall i=1,\dots,n$ by definition. The same can be concluded for every $c_i=d$.

Finally, we note that if $s=1$ then $\max_{i=1,\dots,n} c_i\leq 1$ which corresponds to require $(c_1,\dots,c_n)\in[d,1]^{n}$. $\square$

\section{Price of stability}
\label{sec:price}

In Section \ref{sec:game},we have seen that our game may have at most two Nash equilibria. Those equilibria represent the outcome of the strategic interaction of the players, i.e. the advisor and the customers (the individual investors), to maximize their own utilities. However, their decisions may  differ from what could be achieved if the overall maximum utility would be sought. Therefore, the utility achieved under a Nash equilibrium could be globally not efficient. In the cases with a single Nash equilibrium, this loss of efficiency can be computed through the Price of Anarchy. In the case of more Nash equilibria, like ours, that concept has been generalized into the Price of Stability (PoS) \cite{Anshelevich}. In this section, we compute the Price of Stability for our game.

For the price of stability we adopt the definition of \cite{Anshelevich}:
\begin{equation}
\label{eq:pos}
    \text{PoS} = \frac{\text{value of best equilibrium}}{\text{value of optimal solution}}\, .
\end{equation}

Let us denote $u_A(\textbf{c},s,w,x):=u_A(\textbf{c},s)$ and $u_{CL_i}(c_i,d_i,s):=u_{CL_i}(c_i)$. We now calculate the utility functions outcomes in Nash equilibria $P^\ast$ and $P^\dagger$. Then:
\begin{align}
\label{eq:utility(Past)}
u_A(P^\ast)&=-\alpha \left(a-x\right)^2-\beta n\left(w-\frac{1}{2\zeta}\frac{r_s-r_{d}}{a -d}- a\right)^2- \frac{\gamma n}{4\zeta^2}\left(\frac{r_s-r_{d}}{a -d}\right)^2\notag \\
u_{CL_i}(P^\ast)&=r_s+\frac{1}{4\zeta} \left(\frac{r_s-r_{d}}{a -d}\right)^2
\end{align}
and
\begin{align}
\label{eq:utility(Pdagger)}
u_A(P^\dagger)&=-\alpha \left(b-x\right)^2-\beta n\left(w-\frac{1}{2\zeta}\frac{r_s-r_{d}}{b -d}- b\right)^2-\frac{\gamma n}{4\zeta^2}\left(\frac{r_s-r_{d}}{b -d}\right)^2\notag \\
u_{CL_i}(P^\dagger)&=r_{s}+\frac{1}{4\zeta} \left(\frac{r_s-r_{d}}{b -d}\right)^2\, .
\end{align}
The social welfare, i.e. the total utility of the agents, is:
\begin{align}
\label{eq:SW}
SW(\textbf{c},s)&=u_A (\textbf{c},s) +\sum_{i=1}^n u_{CL_i}(c_i,s) \notag \\
&=-\alpha \left(s-x\right)^2-\beta \sum_{i=1}^n\left(w-c_i\right)^2-(\gamma+\zeta) \sum_{i=1}^n\left(s-c_i\right)^2 \notag  \\
&+ r_d n+  \frac{r_s-r_{d}}{s-d} \left[-dn+ \sum_{i=1}^n c_i\right]\, .
\end{align}
Because of the mixed terms in $c_i$ and $s$, the optimal solution of $i$-th customer depends on the choices done by financial advisor.

Whether the maxima of $SW$ belong to $\mathcal{D}$ or $\partial\mathcal{D}$, is a question that is addressed and fully solved by Proposition \ref{p:1} below.

In the following, for any complex number $z=x+iy$ where $x$ and $y$ are real numbers, the absolute value or modulus of $z$ is denoted $|z|$ and is defined by $|z|=\sqrt {x^{2}+y^{2}}$.

The following preliminary result concerns the roots of a quartic equation. A general method for solving quartic equations is found in Cardano's Ars Magna, but it is attributed to Cardano's assistant Ludovico Ferrari (1522-1565) \cite{leung1992polynomials}.
\begin{proposition}
\label{p:1}
Let us consider
\begin{equation}
\label{eq:pol}
\omega_4 z^4+\omega_3 z^3+\omega_1 z+\omega_0=0\, ,
\end{equation}
where $\omega_0,\omega_4>0$, $\omega_1,\omega_3\in\mathbb R$ ($\omega_1,\omega_3$ both negative or both positive). Let also
\begin{align}
\label{eq:quant_pol}
\Delta&=256\omega_4^{3}\omega_0^{3}-192\omega_4^{2}\omega_3\omega_1\omega_0^{2}-27\omega_4^{2}\omega_1^{4}-6\omega_4\omega_3^{2}\omega_1^{2}\omega_0-27\omega_3^{4}\omega_0^{2}-4\omega_3^{3}\omega_1^{3}\notag \\
D&=64\omega_4^{3}\omega_0-16\omega_4^{2}\omega_3\omega_1-3\omega_3^{4}\, , \quad P=-3\omega_3^2\, , \quad R=\omega_3^3+8\omega_1\omega_4^2\, .
\end{align}
The following are proved:
\begin{itemize}
    \item[(i)] All the roots of (\ref{eq:pol}) are not-real if and only if $\Delta>0$ and $D> 0$.
   \item[(ii)] There exists at least one root of (\ref{eq:pol}) which has positive real part.
  \item[(iii)] Let 
  \begin{align}
    \label{eq:Omega_set}
    \Omega=\Bigl\{\omega_0,\omega_1,&\omega_3,\omega_4\, :\, \omega_4-|\omega_1|-|\omega_3|+\omega_0>0\, , \notag \\ 
    & 4\omega_4-|\omega_1|-3|\omega_3|<0\, ,  \, (\Delta\leq 0\ \text{or}\ D\leq 0)\Bigr\}
  \end{align}
  be a subset of the admissible parameter region $\mathcal{R}$. Then, $\forall \omega_i\in\Omega$ all the roots of (\ref{eq:pol}) have modulus $> 1$.
\end{itemize}
\end{proposition}
\proof
The completed proof is given in Appendix. $\square$

We now turn our attention to finding the maximum of the function $(\textbf{c},s)\to SW(\textbf{c},s)$ in the set $\mathcal{D}$ described in (\ref{eq:domain1}).

\begin{theorem}\label{th:1}
Let $\Delta$, D, P, R and $\Omega$ like in Proposition \ref{p:1}. Let also SW be the social welfare function as in (\ref{eq:SW}) and let
\begin{align}
\label{eq:coeff_y}
\omega_0&= n (r_d - r_s)^2 \notag \\
\omega_1&= 2 \beta n (r_d - r_s) (d - w) \notag \\
\omega_2&=0 \notag\\
\omega_3&= 4 \left[\beta n (d - w) (\gamma + \zeta) + \alpha (d - x) (\beta + \gamma + \zeta)\right] \notag \\
\omega_4&= 4 \left[\beta n (\gamma + \zeta) + \alpha (\beta + \gamma + \zeta)\right]\, .
\end{align}
Then the following claims hold.
\begin{itemize}
    \item[i)] If $\Delta>0$ and $D> 0$, or $\omega_i\in\Omega$ $\forall i=0,\dots,4$, the function SW attains its maximum in a point belonging to $\partial\mathcal{D}$.
    \item[ii)] Let $y=s-d\in[0,1]$. In all the other cases in which SW results concave, the function attains its maximum in a point $(c_1,\dots,c_n,s)\in\mathcal{D}$ such that
\begin{equation}
\label{eq:poly_y}
\omega_4 y^4+\omega_3 y^3+\omega_2 y^2+\omega_1 y+\omega_0=0
\end{equation}
and
\begin{equation}
c_1=\dots=c_n=\frac{2\beta w+2(\gamma+\zeta) s   +\frac{r_s-r_{d}}{s-d}}{2\beta +2(\gamma+\zeta)}\, .
\end{equation}
\end{itemize}
\end{theorem}
\proof
Let us first consider the maximum points of $SW$ that are internal to $\mathcal{D}$, namely in
\begin{equation}
    \operatorname{int}\left(\mathcal{D}\right):=\left\{(c_1,\dots,c_n,s)\, \big|\, d < c_i< s\, \forall i=1,\dots,n,\, s\in[d,1]\right\}\, .
\end{equation}
Being $SW$ of class $C^\infty$, the maximum points in $\operatorname{int}\left(\mathcal{D}\right)$ can be found amongst the stationary points, in other words amongst the points $(c_1,\dots,c_n,s)\in \operatorname{int}\left(\mathcal{D}\right)$ such that $\nabla SW(c_1,\dots,c_n,s) = (0,\dots,0)$. We have that
\begin{equation}
    \begin{cases}
        \frac{\partial SW(c_1,\dots,c_n,s)}{\partial c_1}=0 \\  \dots \\ \frac{\partial SW(c_1,\dots,c_n,s)}{\partial c_n}=0\\ \frac{\partial SW(c_1,\dots,c_n,s)}{\partial s}=0\, ,
    \end{cases}
\end{equation}
thus
\begin{equation}
c_1=\dots=c_n=\frac{2\beta w+2(\gamma+\zeta) s   +\frac{r_s-r_{d}}{s-d}}{2\beta +2(\gamma+\zeta)}
\end{equation}
and
\begin{align}
&-2\alpha \left(s-x\right)-2(\gamma+\zeta)n \left(s- \frac{2\beta w+2(\gamma+\zeta) s   +\frac{r_s-r_{d}}{s-d}}{2\beta +2(\gamma+\zeta)}\right)+\notag \\
&- \frac{r_s-r_{d}}{(s-d)^2} \left[  \frac{2\beta (w-d)+2(\gamma+\zeta) (s-d)   +\frac{r_s-r_{d}}{s-d}}{2\beta +2(\gamma+\zeta)} \right]n=0
\end{align}
Rearrange the terms of the latter equation:
\begin{align}
&2\alpha \left(s-x\right)(2\beta +2(\gamma+\zeta)) (s-d)^3+\notag \\
&+2(\gamma+\zeta)n \left[2\beta (s- w)(s-d)^3-(r_s-r_{d})(s-d)^2\right]+\notag\\
&+ (r_s-r_{d}) \left[ 2\beta (w-d) (s-d)+2(\gamma+\zeta) (s-d)^2   +(r_s-r_{d}) \right]n=0
\end{align}
which, substituting the new variable $y=s-d\in[0,1]$, yields the polynomial equation (\ref{eq:poly_y}), where $\omega_0$, ..., $\omega_4$ are described in (\ref{eq:coeff_y}). This proves claim ii).

However, as we can see, $\omega_0$ and $\omega_4$ are $\geq0$. Accordingly, by assumption of claim i) and from Proposition \ref{p:1} the social welfare function does not assume (admissible) maxima in $\mathcal{D}$ and so we have to focus on $\partial\mathcal{D}$. And this proves claim i). $\square$

According to Theorem \ref{th:1}, let us denote the maximum values assumed by function SW with $\text{SW}_M$ and assume that it is global. Then, from the definition of PoS described by Eq. (\ref{eq:pos}), we have
\begin{equation}
\label{eq:pos2}
    \text{PoS}=\frac{\max\left\{u_A(P^\ast) +\sum_{i=1}^n u_{CL_i}(P^\ast)\, , \, u_A(P^\dagger) +\sum_{i=1}^n u_{CL_i}(P^\dagger)\right\}}{\text{SW}_M}
\end{equation}
where the utility functions estimated in $P^\ast$ and $P^\dagger$ are shown in (\ref{eq:utility(Past)}) and (\ref{eq:utility(Pdagger)}). One can look at Eq. (\ref{eq:pos2}) as how much the central authority can earn if he/she can intervene in the game, helping the players (advisor, customers) converge to a good Nash equilibrium.

\section{Conclusions and further works}

Investors usually resort to financial advisors (paid by a bank) to improve their investment process until the point of complete delegation on investment decisions. Surely, financial advice is potentially a correcting factor in investment decisions but, in the past, the media and regulators blamed biased advisors for manipulating the expectations of naive investors. Then we wondered whether that was indeed the case and we built an ABM for the communication process between bank, advisors and investors. 

We defined a compromise for the financial advisor (between a sufficient reward by bank and to keep his/her reputation), and a compromise for the customers (between the desired return and the proposed return by advisor). In this way, the notion of PoS --- which we also analytically formulated --- naturally arisen in our model.

Moreover, we obtained two Nash equilibria and the best response functions of the resulting game. Anyway, one of these equilibria is not always acceptable:
\begin{itemize}
    \item The presence of a very truthful advisor translates into the presence of only one Nash equilibria (represented by his/her internal opinion). This case corresponds to an advisor that is very sensitive to the difference between his/her stated and true opinion, i.e. to the difference between what he/she says and what he/she really thinks.
    \item If the advisor's desire to influence the customers is irrelevant, only one equilibria survives and his/her equilibrium is represented by his/her internal opinion.
    \item The same equilibria associated to advisor's internal opinion survives when the sensitivity of customer to cognitive dissonance becomes strong. Cognitive dissonance provides customers with an incentive to modify their behavior to reduce the ``cost'' of the lack of agreement between advisor and customers. Then this case corresponds to customers that are very sensitive to the difference between his/her opinion and advisor's stated opinion.
\end{itemize}
Then, by describing the parameter regions in which both equilibria result acceptable, we shown that they exist whether customers expect a return bigger than what advisor proposes instead to them: we can say that greediness/naivety of the customers emerge naturally from the model. 

The results of the paper concern the special case of homogeneous investors. It would be interesting to extend the results of the paper to the more general case of \emph{non-homogeneous investors}. Moreover, although a number of analytic results obtained here for this special case, a closed-form expression of PoS seems to be unmanageable. We feel that a widespread use of simulation tools may help the understanding these open questions in future works.

\appendix
\renewcommand{\thesection}{\Alph{section}}
\section{Appendix}

\subsection{Proof of Proposition \ref{p:1}}
According to the theory of quartic equations \cite{tignol2015galois}, all the roots of (\ref{eq:pol}) are non-real only in the following cases
\begin{itemize}
    \item $\Delta> 0$ and $D> 0$
    \item $\Delta> 0$ and $P> 0$
    \item $\Delta=0$ and $D=0$ and $P> 0$ and $R=0$.
\end{itemize}
Since $P=-3\omega_3^2< 0$, we see that all the roots of (\ref{eq:pol}) are non-real if and only if $\Delta>0$ and $D> 0$.

For the proof of (ii) let $f(z)$ be LHS of (\ref{eq:pol}). Then, $f'(z)=4\omega_4z^3+3\omega_3z^2+\omega_1$. We have that $f'(z)=0$ has only one real root because the discriminant of the cubic equation $f'(z)=0$, $\Delta_3$ (see \cite{tignol2015galois}), is negative
\begin{equation}
\Delta_3=-108\omega_3^3\omega_1-432\omega_4^2\omega_1^2< 0    \, .
\end{equation}
It follows that the number of the real roots of the quartic equation (\ref{eq:pol}) is at most two.

Our proof proceeds by reductio ad absurdum. Let us assume that all the roots of (\ref{eq:pol}) have negative real part. The roots may be $a$, $b$, $c+di$, $c-di$ where $a$, $b$, $c$, $d\in\mathbb R^+$ with $a< 0$, $b< 0$, $c< 0$, $d> 0$ or $a+bi$, $a-bi$, $c+di$, $c-di$ where $a$, $b$, $c$, $d\in\mathbb R^+$ with $a< 0$, $b> 0$, $c< 0$, $d> 0$.

Since the coefficient of $z^2$ is $0$, we get, by Vieta's formulas (see ``Newton's Identities'' in \cite{borwein2012polynomials}), 
\begin{equation}
    \frac{0}{\omega_4}=ab+2ac+2bc+c^2+d^2
\end{equation}
for the first case and
\begin{equation}
    \frac{0}{\omega_4}=a^2+b^2+4ac+c^2+d^2
\end{equation}
for the second. In both cases, the LHS equals $0$ while the RHS is positive. This is impossible.

For the proof of (iii) we already know, from (ii), that $f'(z)=0$ has only one real root. Besides, we have $f''(z)=0$ $\iff$ $z=-\frac{\omega_3}{2\omega_4}$ and $z=0$.

Now, depending on the sign of $\omega_1$ and $\omega_3$ we have two different cases.

\begin{figure}[ht]
\begin{center}
\begin{tikzpicture}[scale=0.7]
\begin{axis}[axis x line=middle, axis y line=middle,xlabel={$z$}, ylabel={$y$}, axis equal, minor tick num=1]
\addplot[domain=-2:2]{(x^3-x+2)/8};
\filldraw (-1.5,0) circle[radius=0.7pt];
\node[below right=0.5pt of {(-1.5,0)}, outer sep=0.1pt,fill=white] {$\alpha$};
\draw[dashed] (-1,-0.035) -- (-1,0.3);
\end{axis}
\end{tikzpicture}
\quad
\begin{tikzpicture}[scale=0.7]
\begin{axis}[axis x line=middle, axis y line=middle,xlabel={$x$}, ylabel={$y$}, axis equal, minor tick num=1]
\addplot[domain=-3:0.3]{(6*(x+0.5)^2+19*(x+0.5)+14)/10-0.3};
\draw[dashed] (-1,-0.035) -- (-1,0.3);
\end{axis}
\end{tikzpicture}
\end{center}
\caption{On the left: a graph of $f'(z)$. On the right: a graph of $f(z)$. As we can see, $f'(\alpha)=0$, $f'(-1)>0$, $f'(0)>0$ and $f(-1)>0$.}
\label{fig:1}
\end{figure}
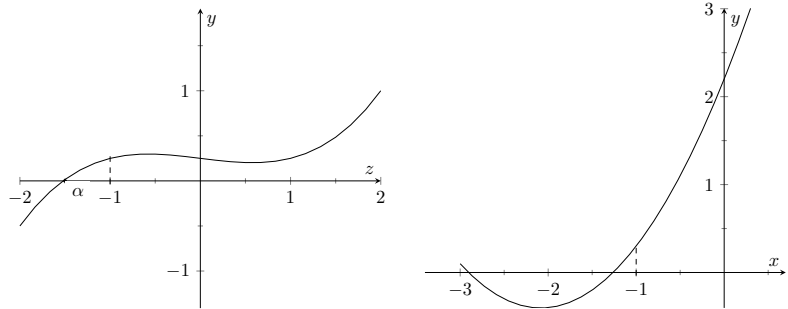

\textbf{Case 1}. Let $\omega_1> 0$ and $\omega_3> 0$. Since $f'(0)=\omega_1> 0$, we see that $f'(z)=0$ has only one real root $z=\alpha$ where $\alpha< 0$. It is necessary that $f(-1)=\omega_4-\omega_3-\omega_1+\omega_0> 0$ and that $f(z)=0$ has at least one real root, i.e. $\Delta\leq 0$ or $D\leq 0$ from (i). Since we have $-\frac{\omega_3}{2\omega_4}< 0$ and $f'(0)=\omega_1> 0$ considering graphs in Fig. \ref{fig:1}, we see that it is necessary that $f'(-1)=-4\omega_4+3\omega_3+\omega_1> 0$. 

On the other hand, if $f(-1)> 0$, $f'(-1)> 0$ and $(\Delta\leq 0$ or $D\leq 0)$ then, we see that all the real roots of (\ref{eq:pol}) have modulus greater than 1.

\begin{figure}[ht]
\begin{center}
\begin{tikzpicture}[scale=0.7]
\begin{axis}[axis x line=middle, axis y line=middle,xlabel={$z$}, ylabel={$y$}, axis equal, minor tick num=1]
\addplot[domain=-2:2.2]{(x^3-x+2)/8-0.7};
\filldraw (1.74,0) circle[radius=0.7pt];
\node[below right=2.3pt of {(1.74,0)}, outer sep=0.1pt,fill=white] {$\beta$};
\draw[dashed] (1,-0.27) -- (1,-0.4);
\end{axis}
\end{tikzpicture}
\quad
\begin{tikzpicture}[scale=0.7]
\begin{axis}[axis x line=middle, axis y line=middle,xlabel={$x$}, ylabel={$y$}, axis equal, minor tick num=1]
\addplot[domain=-0.1:4]{(6*(x-4)^2+19*(x-4)+14)/10-0.3};
\draw[dashed] (1,-0.035) -- (1,0.8);
\end{axis}
\end{tikzpicture}
\end{center}
\caption{On the left: a graph of $f'(z)$. On the right: a graph of $f(z)$. As we can see, $f'(\beta)=0$, $f'(0)<0$, $f'(1)<0$ and $f(1)>0$.}
\label{fig:2}
\end{figure}
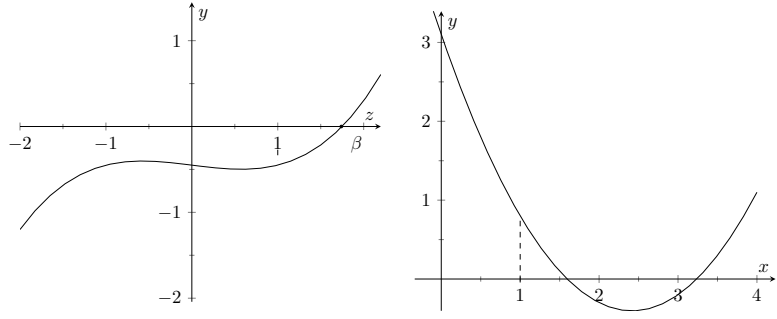

\textbf{Case 2}. Let $\omega_1< 0$ and $\omega_3< 0$. Since $f'(0)=\omega_1< 0$, we see that $f'(z)=0$ has only one real root $z=\beta$ where $\beta> 0$. It is necessary that $f(1)=\omega_4+\omega_3+\omega_1+\omega_0> 0$ and that $f(z)=0$ has at least one real root, i.e. $\Delta\leq 0$ or $D\leq 0$ from (i). Since we have $0< -\frac{\omega_3}{2\omega_4}$ and $f'(0)=\omega_1< 0$ considering graphs in Fig. \ref{fig:2}, we see that it is necessary that $f'(1)=4\omega_4+3\omega_3+\omega_1< 0$. 

On the other hand, if  $f(1)> 0$, $f'(1)< 0$ and ($\Delta\leq 0$ or $D\leq 0$) then, we see that all the real roots of (\ref{eq:pol}) have modulus greater than $1$.

From the two cases, all the real roots of (\ref{eq:pol}) have modulus $> 1$ whenever $\omega_0$, $\dots$, $\omega_4$ belong to the subset $\Omega$. $\square$

\subsection{Proof of Proposition \ref{p:pv}}
Let's start to prove (\ref{eq:adm_ast}) and focus on (\ref{eq:domain1}). Then the admissible parameter region in which $P^\ast$ is an acceptable Nash equilibrium derives from the following inequalities:
\begin{equation}
    0\leq d\leq \frac{1}{2\zeta}\frac{r_s-r_{d}}{a -d}+ a \leq a\leq 1
\end{equation}
from which
\begin{equation}
\label{eq:adm_Past}
\begin{cases}
    a\leq 1\\
    a\geq 0 \\
    d\leq a\\
    d\leq \frac{1}{2\zeta}\frac{r_s-r_{d}}{a -d}+ a \\
    \frac{1}{2\zeta}\frac{r_s-r_{d}}{a -d}+ a \leq a
\end{cases} \quad \Rightarrow \quad 
\begin{cases}
    a\leq 1\\
    a\geq 0 \\
    d\leq a\\
    d-a \leq \frac{1}{2\zeta}\frac{r_s-r_{d}}{a -d} \\
    \frac{1}{2\zeta}\frac{r_s-r_{d}}{a -d} \leq 0
\end{cases}
\end{equation}
By third equation of (\ref{eq:adm_Past}) --- $d\leq a$ --- from $a\geq0$ and, from $a$'s definition in (\ref{eq:a-b}), the system (\ref{eq:adm_Past}) becomes
\begin{numcases}{}
    \frac{d+  x}{2} + \frac{1}{2} \sqrt{\left(  d-  x \right)^2  +\frac{2\gamma n}{\alpha\zeta} (r_s-r_{d}) }\leq 1 \label{eq:adm_Past_2-a}\\
    d\leq \frac{d+  x}{2} + \frac{1}{2} \sqrt{\left(  d-  x \right)^2  +\frac{2\gamma n}{\alpha\zeta} (r_s-r_{d}) } \label{eq:adm_Past_2-b}\\
        r_d-2\zeta\left(\frac{x-  d}{2} + \frac{1}{2} \sqrt{\left(  d-  x \right)^2  +\frac{2\gamma n}{\alpha\zeta} (r_s-r_{d}) }\right)^2\leq r_s< r_{d} \label{eq:adm_Past_2-c}
\end{numcases}
Let us consider inequality (\ref{eq:adm_Past_2-a}):
\begin{equation}
        \sqrt{\left(  d-  x \right)^2  +\frac{2\gamma n}{\alpha\zeta} (r_s-r_{d}) }\leq 2-d-x\, .
\end{equation}
It has solution:
\begin{equation}
    \begin{cases}
        \left(  d-  x \right)^2  +\frac{2\gamma n}{\alpha\zeta} (r_s-r_{d})\geq 0\\
        d+x\leq 2 \quad (\text{which is checked because }d,x\in[0,1]) \\
         r_s \leq r_{d}+ \frac{2\alpha\zeta}{\gamma n} (1-d-x+dx)
    \end{cases}
\end{equation}
which can be simplified in
\begin{equation}
\label{eq:adm_Past_2-a-solution}
r_{d} -\frac{\alpha\zeta}{2\gamma n}(x-d)^2 \leq r_s \leq r_{d}+ \frac{2\alpha\zeta}{\gamma n} (1-d)(1-x)\, .
\end{equation}
Let us consider inequality (\ref{eq:adm_Past_2-b}),
\begin{equation}
     \sqrt{\left(  d-  x \right)^2  +\frac{2\gamma n}{\alpha\zeta} (r_s-r_{d}) }\geq d-x\, ,
\end{equation}
whose solution is
\begin{equation}
    \begin{cases}
        \left(  d-  x \right)^2  +\frac{2\gamma n}{\alpha\zeta} (r_s-r_{d}) \geq 0\\
        d-x\geq 0\\
        \left(  d-  x \right)^2  +\frac{2\gamma n}{\alpha\zeta} (r_s-r_{d}) \geq (d-x)^2
    \end{cases}\quad \cup \quad
    \begin{cases}
        \left(  d-  x \right)^2  +\frac{2\gamma n}{\alpha\zeta} (r_s-r_{d}) \geq 0\\
        d-x< 0
    \end{cases}
\end{equation}
i.e.
\begin{equation}
\label{eq:adm_Past_3}
    \begin{cases}
        \left(  d-  x \right)^2  +\frac{2\gamma n}{\alpha\zeta} (r_s-r_{d}) \geq 0\\
        d-x\geq 0\\
        \frac{2\gamma n}{\alpha\zeta} (r_s-r_{d}) \geq 0
    \end{cases}\quad \cup \quad
    \begin{cases}
        \left(  d-  x \right)^2  +\frac{2\gamma n}{\alpha\zeta} (r_s-r_{d}) \geq 0\\
        d-x< 0
    \end{cases}
\end{equation}
But the third inequality of the first system of (\ref{eq:adm_Past_3}) is at odds with (\ref{eq:adm_Past_2-c}) and, hence, the solution of (\ref{eq:adm_Past_2-b}) comes down to
\begin{equation}
\label{eq:adm_Past_2-b-solution}
\begin{cases}
           r_s\geq r_{d} -\frac{\alpha\zeta}{2\gamma n}(d-x)^2\\
        d<x 
    \end{cases}
\end{equation}
Let us consider inequality (\ref{eq:adm_Past_2-c}), which can be rewritten as follows
\begin{align}
\label{eq:obs}
r_d-\frac{\zeta}{2}\Biggl[2\left(  d-  x \right)^2  &+\frac{2\gamma n}{\alpha\zeta} (r_s-r_{d})+\notag\\
&+2(x-d)  \sqrt{\left(  d-  x \right)^2  +\frac{2\gamma n}{\alpha\zeta} (r_s-r_{d}) } \Biggr]    \leq r_s
\end{align}
and, since $d<x$ --- by (\ref{eq:adm_Past_2-b-solution}) ---, it follows that
\begin{equation}
      \sqrt{\left(  d-  x \right)^2  +\frac{2\gamma n}{\alpha\zeta} (r_s-r_{d}) } \geq -\left(\frac{1}{\zeta} + \frac{\gamma n}{\zeta\alpha}\right) \frac{r_s-r_{d}}{x-d} - \left(  x-d \right)
\end{equation}
from which
\begin{equation}
\label{eq:adm_Past_6}
      \begin{cases}
        \left(  d-  x \right)^2  +\frac{2\gamma n}{\alpha\zeta} (r_s-r_{d}) \geq 0\\
        -\left(\frac{1}{\zeta} + \frac{\gamma n}{\zeta\alpha}\right) \frac{r_s-r_{d}}{x-d} - \left(  x-d \right)\geq 0\\
        \left(  d-  x \right)^2  +\frac{2\gamma n}{\alpha\zeta} (r_s-r_{d}) \geq \left(\left(\frac{1}{\zeta} + \frac{\gamma n}{\zeta\alpha}\right) \frac{r_s-r_{d}}{x-d} + \left(  x-d \right)\right)^2
    \end{cases}
\end{equation}
\begin{equation}
\label{eq:adm_Past_7}
    \cup
    \begin{cases}
        \left(  d-  x \right)^2  +\frac{2\gamma n}{\alpha\zeta} (r_s-r_{d}) \geq 0\\
        -\left(\frac{1}{\zeta} + \frac{\gamma n}{\zeta\alpha}\right) \frac{r_s-r_{d}}{x-d} - \left(  x-d \right)<0
    \end{cases}
\end{equation}
It should be noted that $d\neq x$. Actually, if we substituted $d=x$ in Eq. (\ref{eq:obs}), we would get $r_d-r_s+\frac{\gamma n}{\alpha} (r_d-r_{s}) \leq 0$ which does not have solutions for $r_s<r_d$.

We can rewrite system (\ref{eq:adm_Past_6}) in the following way
\begin{equation}
          \begin{cases}
        r_s\geq r_{d} -\frac{\alpha\zeta}{2\gamma n}(x-d)^2\\
       r_s\leq r_{d} - \frac{\zeta\alpha}{\alpha + \gamma n} \left(  x-d \right)^2 \\
        \left(  d-  x \right)^2  +\frac{2\gamma n}{\alpha\zeta} (r_s-r_{d}) \geq \left(\left(\frac{1}{\zeta} + \frac{\gamma n}{\zeta\alpha}\right) \frac{r_s-r_{d}}{x-d} + \left(  x-d \right)\right)^2
    \end{cases}
\end{equation}
The solutions of the first two inequalities intersect if $\alpha> \gamma n$ while for the last one we have
\begin{equation}
    \frac{2\gamma n}{\alpha\zeta} (r_s-r_{d}) \geq \left(\frac{1}{\zeta} + \frac{\gamma n}{\zeta\alpha}\right)^2 \left(\frac{r_s-r_{d}}{x-d}\right)^2 + 2\left(\frac{1}{\zeta} + \frac{\gamma n}{\zeta\alpha}\right) (r_s-r_{d}) 
\end{equation}
which becomes
\begin{equation}
     0 \geq \left(\frac{1}{\zeta} + \frac{\gamma n}{\zeta\alpha}\right)^2 \left(\frac{r_s-r_{d}}{x-d}\right)^2 + \frac{2}{\zeta} (r_s-r_{d}) \, ,
\end{equation}
whose solution is
\begin{equation}
r_d-2\zeta \alpha^2 \left(\frac{x-d}{\alpha+\gamma n}\right)^2    \leq r_s< r_d
\end{equation}
Then system (\ref{eq:adm_Past_6}) has an empty solution if $\alpha < \gamma n$ and becomes
\begin{equation}
          \begin{cases}
        r_{d} -\frac{\alpha\zeta}{2\gamma n}(x-d)^2 \leq  r_s\leq r_{d} - \frac{\zeta\alpha}{\alpha + \gamma n} \left(  x-d \right)^2 \\
        r_d-2\zeta \alpha^2 \left(\frac{x-d}{\alpha+\gamma n}\right)^2    \leq r_s< r_d
    \end{cases}
\end{equation}
for $\alpha> \gamma n$. The solution of (\ref{eq:adm_Past_6}) is empty for $\alpha < \gamma n$ but, for $\alpha> \gamma n$ it is 
\begin{equation}
\label{eq:adm_Past_8}
    r_d-2\zeta \alpha^2 \left(\frac{x-d}{\alpha+\gamma n}\right)^2 \leq r_s \leq r_{d} - \frac{\zeta\alpha}{\alpha + \gamma n} \left(  x-d \right)^2
\end{equation}
since we have that $r_{d} -\frac{\alpha\zeta}{2\gamma n}(x-d)^2\leq r_d-2\zeta \alpha^2 \left(\frac{x-d}{\alpha+\gamma n}\right)^2$ for each parameters' values. We now focus on system (\ref{eq:adm_Past_7}):
\begin{equation}
     \begin{cases}
        r_s\geq r_{d} -\frac{\alpha\zeta}{2\gamma n}(x-d)^2\\
        \left(\frac{1}{\zeta} + \frac{\gamma n}{\zeta\alpha}\right) \frac{r_s-r_{d}}{x-d} + \left(  x-d \right)>0
    \end{cases} \quad \Rightarrow \quad      
    \begin{cases}
        r_s\geq r_{d} -\frac{\alpha\zeta}{2\gamma n}(x-d)^2\\
        r_s> r_{d} - \frac{\zeta\alpha}{\alpha + \gamma n} \left(  x-d \right)^2
    \end{cases}
\end{equation}
whose solution is
\begin{equation}
\label{eq:adm_Past_9}
    \left\{ r_s\geq r_{d} -\frac{\alpha\zeta}{2\gamma n}(x-d)^2\, , \, \alpha< \gamma n\right\} \cup \left\{ r_s> r_{d} - \frac{\zeta\alpha}{\alpha + \gamma n} \left(  x-d \right)^2 \, , \, \alpha> \gamma n \right\}\, .
\end{equation}
Accordingly, by merging (\ref{eq:adm_Past_8}) --- for $\alpha>\gamma n$ --- and (\ref{eq:adm_Past_9}) we obtain the solution of (\ref{eq:adm_Past_2-c}):
\begin{align}
    &\left\{ r_s\geq r_{d} -\frac{\alpha\zeta}{2\gamma n}(x-d)^2\, , \, \alpha< \gamma n\right\} \cup \left\{ r_s> r_{d} - \frac{\zeta\alpha}{\alpha + \gamma n} \left(  x-d \right)^2 \, \cup \right. \notag\\
    &\left.\, r_d-2\zeta \alpha^2 \left(\frac{x-d}{\alpha+\gamma n}\right)^2 \leq r_s \leq r_{d} - \frac{\zeta\alpha}{\alpha + \gamma n} \left(  x-d \right)^2 \, , \, \alpha> \gamma n \right\}\, .
\end{align}
Hence, the solution of (\ref{eq:adm_Past_2-c}) is
\begin{equation}
\label{eq:adm_Past_2-c-solution}
\left\{ r_s\geq r_{d} -\frac{\alpha\zeta}{2\gamma n}(x-d)^2\, , \, \alpha< \gamma n\right\} \cup \left\{ r_s \geq r_d-2\zeta \alpha^2 \left(\frac{x-d}{\alpha+\gamma n}\right)^2\, , \, \alpha> \gamma n \right\}
\end{equation}
Let us observe that, by (\ref{eq:adm_Past_2-c}), the system (\ref{eq:adm_Past_2-a})-(\ref{eq:adm_Past_2-c}) admits solution only if $-\frac{1}{2}(x-d)^2\leq 2 (1-d-x+dx)$, which is actually equivalent to $(x+d-2)^2\geq0$. Moreover, $r_d\geq r_{d} -\frac{\alpha\zeta}{2\gamma n}(x-d)^2$. By substituting (\ref{eq:adm_Past_2-a-solution}), (\ref{eq:adm_Past_2-b-solution}) and (\ref{eq:adm_Past_2-c-solution}) in the system (\ref{eq:adm_Past_2-a})-(\ref{eq:adm_Past_2-c}) we obtain
\begin{equation}
\label{eq:adm_Past_11a}
\begin{cases}
    r_{d} -\frac{\alpha\zeta}{2\gamma n}(x-d)^2 \leq r_s \leq r_{d}+ \frac{2\alpha\zeta}{\gamma n} (1-d)(1-x)\\
    d<x\\
    \alpha< \gamma n\\
    r_{d} -\frac{\alpha\zeta}{2\gamma n}(x-d)^2 \leq r_s < r_d
\end{cases} 
\end{equation}
and
\begin{equation}
\label{eq:adm_Past_11b}
\begin{cases}
    r_{d} -\frac{\alpha\zeta}{2\gamma n}(x-d)^2 \leq r_s \leq r_{d}+ \frac{2\alpha\zeta}{\gamma n} (1-d)(1-x)\\
    d<x\\
    \alpha> \gamma n\\
     r_d-2\zeta \alpha^2 \left(\frac{x-d}{\alpha+\gamma n}\right)^2 \leq r_s < r_d
\end{cases}
\end{equation}
Since $d,x\in[0,1]$ we have that, for both systems, $r_{d}+ \frac{2\alpha\zeta}{\gamma n} (1-d)(1-x)\geq r_d$. Moreover, to (\ref{eq:adm_Past_11a}) and (\ref{eq:adm_Past_11b}) should be added the conditions $r_d$, $r_s\in[0,1]$. Hence we obtained thesis (\ref{eq:adm_ast}).

We now come to prove (\ref{eq:adm_ast-copy}), by focusing on (\ref{eq:domain1}). Then the admissible parameter region in which $P^\dagger$ is an acceptable Nash equilibrium derives from the following inequalities:
\begin{equation}
    0\leq d\leq \frac{1}{2\zeta}\frac{r_s-r_{d}}{b -d}+ b \leq b\leq 1
\end{equation}
from which
\begin{equation}
\label{eq:adm_Past-copy}
\begin{cases}
    b\leq 1\\
    b\geq 0 \\
    d\leq b\\
    d\leq \frac{1}{2\zeta}\frac{r_s-r_{d}}{b -d}+ b \\
    \frac{1}{2\zeta}\frac{r_s-r_{d}}{b -d}+ b \leq b
\end{cases} \quad \Rightarrow \quad 
\begin{cases}
    b\leq 1\\
    b\geq 0 \\
    d\leq b\\
    d-b \leq \frac{1}{2\zeta}\frac{r_s-r_{d}}{b -d} \\
    \frac{1}{2\zeta}\frac{r_s-r_{d}}{b -d} \leq 0
\end{cases}
\end{equation}
By third formula of system (\ref{eq:adm_Past-copy}) --- $d\leq b$ --- we can rearrange its fourth and fifth equations as, respectively, $-(b-d)^2 \leq \frac{1}{2\zeta}(r_s-r_{d})$ and $r_s< r_{d}$. Moreover, $b\geq0$ if 
\begin{equation}
    d+  x \geq  \sqrt{\left(  d-  x \right)^2  +\frac{2\gamma n}{\alpha\zeta} (r_s-r_{d}) } \quad \Rightarrow \quad \begin{cases}
    d+x\geq 0\\
    \left(  d-  x \right)^2  +\frac{2\gamma n}{\alpha\zeta} (r_s-r_{d})\geq0 \\
     (d+  x)^2\geq \left(  d-  x \right)^2  +\frac{2\gamma n}{\alpha\zeta} (r_s-r_{d}) 
\end{cases}
\end{equation}
but since $d,x\geq0$ and $r_s< r_{d}$, the latter becomes $r_{d} -\frac{\alpha\zeta}{2\gamma n}(x-d)^2 \leq r_s$. Then the system (\ref{eq:adm_Past-copy}) can be rewritten as
\begin{equation}
\begin{cases}
    b\leq 1\\
    d\leq b\\
    \frac{1}{2\zeta}(r_s-r_{d})+(b-d)^2 \geq 0 \\
     r_{d} -\frac{\alpha\zeta}{2\gamma n}(x-d)^2 \leq r_s< r_{d}
\end{cases}
\end{equation}
or, from $b$'s definition in (\ref{eq:a-b})
\begin{numcases}{}
   \frac{d+  x}{2} - \frac{1}{2} \sqrt{\left(  d-  x \right)^2  +\frac{2\gamma n}{\alpha\zeta} (r_s-r_{d}) }\leq 1    \label{eq:adm_Past_2-copy-a}\\
    d\leq \frac{d+  x}{2} - \frac{1}{2} \sqrt{\left(  d-  x \right)^2  +\frac{2\gamma n}{\alpha\zeta} (r_s-r_{d}) }    \label{eq:adm_Past_2-copy-b}\\
    \frac{1}{2\zeta}(r_s-r_{d})+\left(\frac{x-d}{2} - \frac{1}{2} \sqrt{\left(  d-  x \right)^2  +\frac{2\gamma n}{\alpha\zeta} (r_s-r_{d}) }\right)^2 \geq 0    \label{eq:adm_Past_2-copy-c} \\
    r_{d} -\frac{\alpha\zeta}{2\gamma n}(x-d)^2 \leq r_s< r_{d}  \label{eq:adm_Past_2-copy-d}
\end{numcases}
Let us consider inequality (\ref{eq:adm_Past_2-copy-a}):
\begin{equation}
        \sqrt{\left(  d-  x \right)^2  +\frac{2\gamma n}{\alpha\zeta} (r_s-r_{d}) }\geq d+x-2\, .
\end{equation}
We have
\begin{equation}
    \begin{cases}
        \left(  d-  x \right)^2  +\frac{2\gamma n}{\alpha\zeta} (r_s-r_{d}) \geq 0\\
        d+x-2\geq 0\\
        \left(  d-  x \right)^2  +\frac{2\gamma n}{\alpha\zeta} (r_s-r_{d}) \geq (d+x-2)^2
    \end{cases}\quad \cup \quad
    \begin{cases}
        \left(  d-  x \right)^2  +\frac{2\gamma n}{\alpha\zeta} (r_s-r_{d}) \geq 0\\
        d+x-2\leq0 
    \end{cases}
\end{equation}
The first system does not admit solution because $d,x\in[0,1]$ while the solution of the second one comes down to 
\begin{equation}
\label{eq:adm_Past_2-copy-a-sol}
            \left(  d-  x \right)^2  +\frac{2\gamma n}{\alpha\zeta} (r_s-r_{d}) \geq 0
\end{equation}
which hence is the solution of inequality (\ref{eq:adm_Past_2-copy-a}).

Let us consider inequality (\ref{eq:adm_Past_2-copy-b}),
\begin{equation}
    \sqrt{\left(  d-  x \right)^2  +\frac{2\gamma n}{\alpha\zeta} (r_s-r_{d})}\leq  x-d\, ,
\end{equation}
whose solution is
\begin{equation}
\label{eq:adm_Past_2-copy-b-sol}
\begin{cases}
    x-d\geq 0\\
    \left(  d-  x \right)^2  +\frac{2\gamma n}{\alpha\zeta} (r_s-r_{d})\geq0 \\
     \left(  d-  x \right)^2  +\frac{2\gamma n}{\alpha\zeta} (r_s-r_{d}) \leq (x-d)^2
\end{cases} \Rightarrow \begin{cases}
    x\geq d\\
     r_{d} -\frac{\alpha\zeta}{2\gamma n}(x-d)^2 \leq r_s< r_{d}
\end{cases}
\end{equation}
Let us consider inequality (\ref{eq:adm_Past_2-copy-c}):
\begin{equation}
    \frac{1}{2\zeta}(r_s-r_{d})+\left(\frac{x-  d}{2} - \frac{1}{2} \sqrt{\left(  d-  x \right)^2  +\frac{2\gamma n}{\alpha\zeta} (r_s-r_{d}) }\right)^2\geq 0\, .
\end{equation}
It can be rewritten as follows
\begin{align}
\label{eq:obs-copy}
\frac{r_s-r_{d}}{2\zeta}&+\frac{(x-  d)^2}{4}+\frac{1}{4} \left[\left(  d-  x \right)^2  +\frac{2\gamma n}{\alpha\zeta} (r_s-r_{d})\right]+\notag\\
&-\frac{x-d}{2} \sqrt{\left(  d-  x \right)^2  +\frac{2\gamma n}{\alpha\zeta} (r_s-r_{d}) } \geq0 
\end{align}
and, since $d< x$ --- by (\ref{eq:adm_Past_2-copy-b-sol}) ---, we obtain
\begin{equation}
      \sqrt{\left(  d-  x \right)^2  +\frac{2\gamma n}{\alpha\zeta} (r_s-r_{d}) } \leq x-d +\left(\frac{1}{\zeta} + \frac{\gamma n}{\zeta\alpha}\right) \frac{r_s-r_{d}}{x-d}
\end{equation}
from which
\begin{equation}
\label{eq:adm_Past_6-copy}
      \begin{cases}
        \left(  d-  x \right)^2  +\frac{2\gamma n}{\alpha\zeta} (r_s-r_{d}) \geq 0\\
        x-d+\left(\frac{1}{\zeta} + \frac{\gamma n}{\zeta\alpha}\right) \frac{r_s-r_{d}}{x-d}\geq 0\\
        \left(  d-  x \right)^2  +\frac{2\gamma n}{\alpha\zeta} (r_s-r_{d}) \leq \left(\left(\frac{1}{\zeta} + \frac{\gamma n}{\zeta\alpha}\right) \frac{r_s-r_{d}}{x-d} + \left(  x-d \right)\right)^2
    \end{cases}
\end{equation}
It should be noted that $d\neq x$ because, if we substituted $d=x$ in Eq. (\ref{eq:obs-copy}), we would get $(r_s-r_d)\left[1+\frac{\gamma n}{\alpha}\right] \geq 0$ which does not have solutions for $r_s<r_d$.

We can rewrite system (\ref{eq:adm_Past_6-copy}) in the following way
\begin{equation}
\label{eq:adm_Past_7-copy}
          \begin{cases}
        r_s\geq r_{d} -\frac{\alpha\zeta}{2\gamma n}(x-d)^2\\
       r_s\geq r_{d} - \frac{\zeta\alpha}{\alpha + \gamma n} \left(  x-d \right)^2 \\
        \left(  d-  x \right)^2  +\frac{2\gamma n}{\alpha\zeta} (r_s-r_{d}) \leq \left(\left(\frac{1}{\zeta} + \frac{\gamma n}{\zeta\alpha}\right) \frac{r_s-r_{d}}{x-d} + \left(  x-d \right)\right)^2
    \end{cases}
\end{equation}
For the last inequality we have
\begin{equation}
    \frac{2\gamma n}{\alpha\zeta} (r_s-r_{d}) \leq \left(\frac{1}{\zeta} + \frac{\gamma n}{\zeta\alpha}\right)^2 \left(\frac{r_s-r_{d}}{x-d}\right)^2 + 2\left(\frac{1}{\zeta} + \frac{\gamma n}{\zeta\alpha}\right) (r_s-r_{d}) 
\end{equation}
which becomes
\begin{equation}
     0 \leq \left(\frac{1}{\zeta} + \frac{\gamma n}{\zeta\alpha}\right)^2 \left(\frac{r_s-r_{d}}{x-d}\right)^2 + \frac{2}{\zeta} (r_s-r_{d}) \, ,
\end{equation}
whose (acceptable) solution is
\begin{equation}
r_s\leq r_d-2\zeta \alpha^2 \left(\frac{x-d}{\alpha+\gamma n}\right)^2\, .
\end{equation}
Hence we rewrite system (\ref{eq:adm_Past_7-copy}) as
\begin{equation}
\label{eq:adm_Past_8-copy}
          \begin{cases}
        r_s\geq r_{d} -\frac{\alpha\zeta}{2\gamma n}(x-d)^2\\
       r_s\geq r_{d} - \frac{\zeta\alpha}{\alpha + \gamma n} \left(  x-d \right)^2 \\
        r_s\leq r_d-2\zeta \alpha^2 \left(\frac{x-d}{\alpha+\gamma n}\right)^2
    \end{cases}
\end{equation}
Since
\begin{equation}
    r_{d} - \frac{\zeta\alpha}{\alpha + \gamma n} \left(  x-d \right)^2\leq r_{d} -\frac{\alpha\zeta}{2\gamma n}(x-d)^2\leq r_d-2\zeta \alpha^2 \left(\frac{x-d}{\alpha+\gamma n}\right)^2\leq r_d \, ,
\end{equation}
for $\alpha<\gamma n$, and
\begin{equation}
    r_{d} -\frac{\alpha\zeta}{2\gamma n}(x-d)^2\leq r_d-2\zeta \alpha^2 \left(\frac{x-d}{\alpha+\gamma n}\right)^2\leq r_{d} - \frac{\zeta\alpha}{\alpha + \gamma n} \left(  x-d \right)^2 \leq r_d \, ,
\end{equation}
for $\alpha>\gamma n$, the solution of system (\ref{eq:adm_Past_8-copy}) and then of the inequality (\ref{eq:adm_Past_2-copy-c}), is empty if $\alpha>\gamma n$ but it equals
\begin{align}
\label{eq:adm_Past_2-copy-c-sol}
& r_{d} -\frac{\alpha\zeta}{2\gamma n}(x-d)^2 \leq r_s\leq r_d-2\zeta \alpha^2 \left(\frac{x-d}{\alpha+\gamma n}\right)^2 & \text{if} \ \alpha<\gamma n 
\end{align}
Accordingly, by substituting (\ref{eq:adm_Past_2-copy-a-sol}), (\ref{eq:adm_Past_2-copy-b-sol}) and (\ref{eq:adm_Past_2-copy-c-sol}) into the system (\ref{eq:adm_Past_2-copy-a})-(\ref{eq:adm_Past_2-copy-d}) we obtain
\begin{equation}
\begin{cases}
    d< x\\
    \alpha<\gamma n\\
    r_{d} -\frac{\alpha\zeta}{2\gamma n}(x-d)^2 \leq r_s\leq r_d-2\zeta \alpha^2 \left(\frac{x-d}{\alpha+\gamma n}\right)^2 \\
     r_{d} -\frac{\alpha\zeta}{2\gamma n}(x-d)^2 \leq r_s< r_{d}
\end{cases}
\end{equation}
which turns into
\begin{equation}
\begin{cases}
    d< x\\
    \alpha<\gamma n\\
    r_{d} -\frac{\alpha\zeta}{2\gamma n}(x-d)^2 \leq r_s\leq r_d-2\zeta \alpha^2 \left(\frac{x-d}{\alpha+\gamma n}\right)^2
\end{cases}
\end{equation}
Hence we obtained thesis (\ref{eq:adm_ast-copy}). $\square$

\bibliographystyle{unsrt}  
\bibliography{references}  

\end{document}